%% file: bare_jrnl.tex
\documentclass[10pt,final,journal]{IEEEtran} 

\ifCLASSINFOpdf
 \usepackage[pdftex]{graphicx}
\else
\fi
\usepackage{url}
\usepackage{amsmath}
\usepackage{amssymb}
\usepackage{setspace}
\usepackage{xcolor}
\usepackage[caption=false,font=footnotesize]{subfig}

\DeclareMathOperator\sgn{sgn}
\newcommand{\tran}{^{\footnotesize\mbox{T}}}
\newcommand{\realnumber}{\rm I\!R}
\newcommand*{\matr}[1]{\mathbf{\textbf #1}}

\begin{document}
%
\title{Unified Backstepping Sliding Mode \\ Framework for Airship Control Design}
%
%
%

\author{Henrique~S.~Vieira,
        Ely~C.~de~Paiva,~S{e}rgio~K.~Moriguchi
        and~Jos\'{e} R. H. Carvalho
\thanks{Ely Paiva is with the School of Mechanical Engineering, University of Campinas, Campinas, Brazil, as well as the PhD student Henrique Vieira,  e-mail: (see http://www.fem.unicamp.br/~elypaiva/)}
\thanks{Sergio Moriguchi is with Maua Institute of Technology, SP, and Reginaldo Carvalho is with Inst. of Comput. Univ. Fed. do Amazonas, Manaus, Brazil. }
\thanks{Manuscript received April 19, 2005; revised August 26, 2015.}}

%
%

\markboth{Journal of \LaTeX\ Class Files,~Vol.~14, No.~8, August~2015}%
{Shell \MakeLowercase{\textit{et al.}}: Bare Demo of IEEEtran.cls for IEEE Journals}
%



\maketitle

\begin{abstract}
This paper presents a new kind of vectorial backstepping sliding mode control (BSMC) for the positioning and trajectory tracking of an autonomous robotic airship. Also, a unified framework basis for the design/analysis of vectorial BSMC, as well as sliding mode control (SMC) and backstepping control  (BS) for a system in   “lower triangular block” form is derived. The design framework makes it easier the theoretical-based comparative analysis of performances/robustness between the three nonlinear control approaches. Simulation results for the positioning and tracking of the autonomous airship illustrate the proposal.
\end{abstract}

\begin{IEEEkeywords}
Nonlinear Control, Backstepping, Sliding Mode Control, Airship, Unmanned Aerial Vehicles.
\end{IEEEkeywords}

%
\IEEEpeerreviewmaketitle

\input{introduction.tex}

\input{modeling.tex}

\input{controls.tex}

\input{results.tex}

\input{conclusion.tex}


%



\section*{Acknowledgment}
The authors acknowledge the fundings received from: FAPEAM (p. 253/2014, Dec. 014/2015), CNPq DRONI Project (p. 402112/2013-0), Project INCT-SAC - Autonomous Collaborative Systems - (CNPq 465755/2014-3, FAPESP 2014/50851-0) and Fapesp BEP (p. 2017/11423-0). Also, the authors would like to thank Prof. Luis Rodrigues, from  Concordia University of Montreal, for the discussions on backstepping for UAV flight controllers, as well as the support of Prof. Jos\'{e} R. Azinheira from IST-Portugal.

\ifCLASSOPTIONcaptionsoff
  \newpage
\fi



\bibliographystyle{IEEEtran}
\bibliography{reference}
%



%

\begin{IEEEbiography}[{\includegraphics[width=1in,height=1.25in,clip,keepaspectratio]{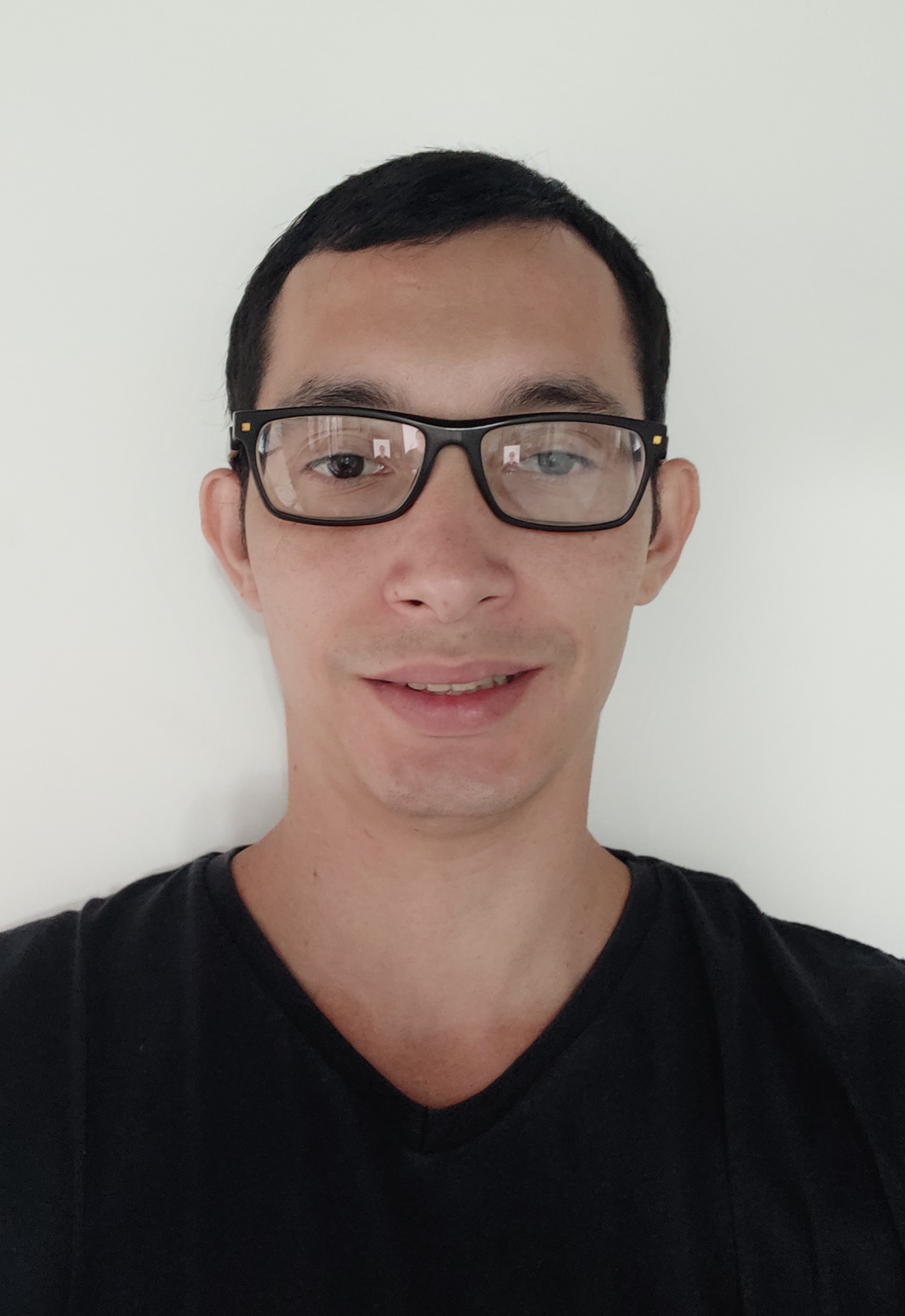}}]{Henrique Vieira} was born in Brazil, in 1989. He has a Bachelor degree in Mechatronics Engineering from Amazonas State University (2012), Master’s in Electrical Engineering from Campinas State University (2015). Currently, he is finishing the PhD in Mechanical Engineering in Campinas State University, in the field of nonlinear control design with applications to an autonomous airship.
\end{IEEEbiography}

\begin{IEEEbiography}
[{\includegraphics[width=1in,height=1.25in,clip,keepaspectratio]{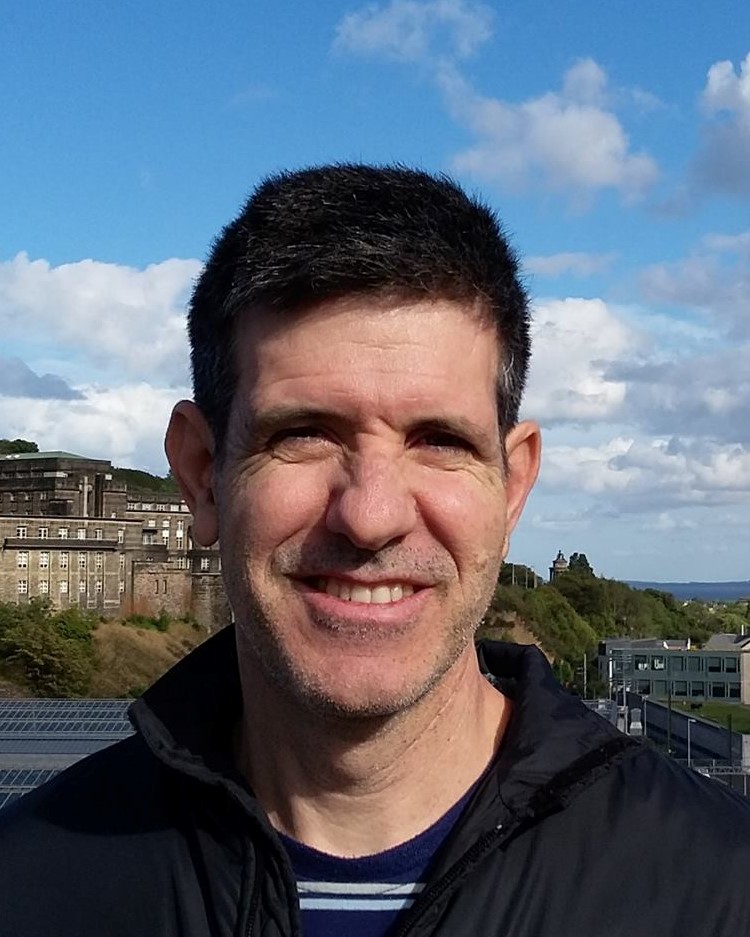}}]{Ely C. de Paiva}
was born in Brazil, in 1965. He
received the Ph.D. degree in Electrical Engineering
(Automation) in the University of Campinas (Unicamp),
Brazil, in 1997. From 1997 until 2009, he
worked as researcher of the Center
for Information Technology Renato Archer, Campinas, Brazil,
within the Robotics and Computer Vision Division.
Since 2010, he is professor of the School of
Mechanical Engineering of Unicamp. In 2018/2019 he was in  research leave at Concordia University of Montreal, working with Optimal Guidance and Motion Planning for unmanned aerial vehicles (UAVs). His research interests include robust and nonlinear
control, mobile robotics, autonomous vehicles,  modelling and flight control,
\end{IEEEbiography}

\begin{IEEEbiography}
[{\includegraphics[width=1in,height=1.25in,clip,keepaspectratio]{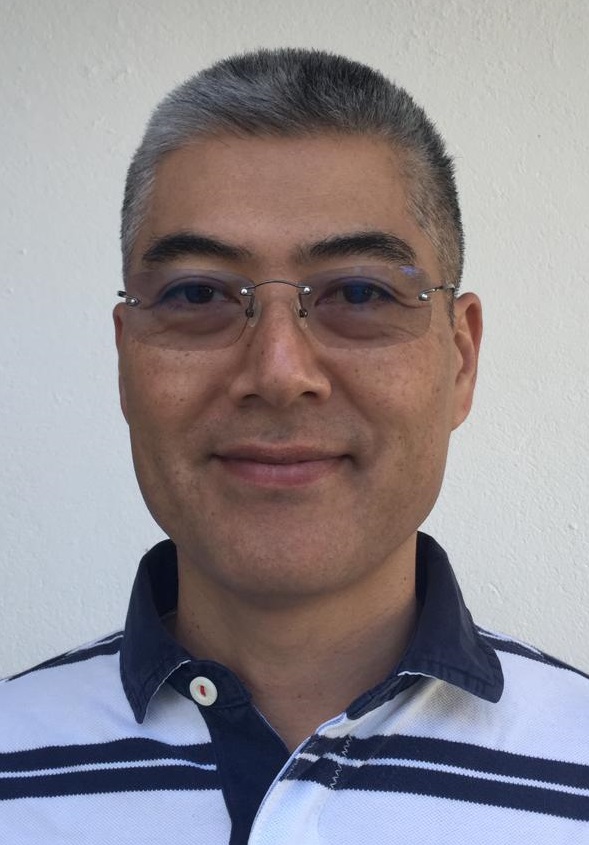}}]{Sergio K. Moriguchi}
 was born in Brazil, in 1965.
He received the Ph.D. degree in Mechanical Engineering in the University of Campinas (Unicamp), Brazil, in 2017.
Since 1991, he is professor of the Instituto Maua de Tecnologia, Sao Paulo, Brazil, at the mechanical department.
His research interests include robust and nonlinear control, flight control of airships, terrestrial vehicles and PLM systems.
\end{IEEEbiography}

\begin{IEEEbiography}
[{\includegraphics[width=1in,height=1.25in,clip,keepaspectratio]{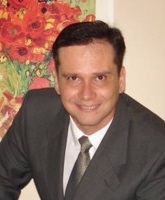}}]
{Jose R. H. Carvalho}
is Doctor in System Engineering by the State University of Campinas on 1997. Dr. Carvalho has up to 20 years of experience on R$\&$D programs. From 1997 to 2001 he was with CTI Renato Archer, a public RD institute, working on robotics and visual servoing. On 2001, he moved to the private sector, working for companies such as Siemens, Genius/Gradient and Intera/Digitron in the area of telecommunication and consumer electronics. During this period, Dr. Carvalho participated in the creation and establishment of three RD departments. On 2009, Dr. Carvalho joined the Institute of Computing of Federal University of Amazonas (UFAM). Since then, Dr. Carvalho has had the opportunity to lead several projects related to field and service robotics, aerial vehicles, and environmental monitoring. Currently, his research interests are focused on intelligent systems applied to field and service robotics.
\end{IEEEbiography}

\end{document}

%% file: introduction.tex
\section{Introduction}
\label{sec:Intro}	

\IEEEPARstart{T}{he} development of unmanned aerial vehicles (UAVs), also known as aerial robotic vehicles,
constitutes an emergent area of scientific and technological research, with a rapid growth of applications in different fields. One of them is
related to agricultural/climatological/environmental monitoring, where the UAV should flight at low altitude and low speed airborne for data gathering \cite{bueno2002project}. For this kind of task, the unmanned airships are the best suited UAVs due to: (i) low interference, including low noise generation, when electrical motors are used; (ii)
hovering capability, with vertical take-off and landing (VTOL), and finally (iii)  greater flight endurance due to the aerostatic lift \cite{bueno2002project,carvalho2014sistemas}.

In this sense, this research is related to the project of an unmanned autonomous airship - NOAMAY Project - that aims the deployment/operation of an airship pilot experiment for surveillance/monitoring in brazilian Amazon forest.\footnote{\url{http://revistapesquisa.fapesp.br/en/2018/09/11/blimp-over-the-forest}}
The NOAMAY prototype is a 11m long airship (Figure 1), an evolution 
of the pioneer  airship\footnote{\url{http://revistapesquisa.fapesp.br/en/2003/02/01/intelligence-and-pilotless}} of Project AURORA \cite{bueno2002project,depaiva2006,moutinho2016airship},
using an innovative configuration of four electrical
tilting propellers, that improves controllability and
maneuverability \cite{vieira2017controle,Vie:19}. In this context, this paper is related to the development of three different nonlinear control approaches, designed under a common unified framework, to be used in NOAMAY airship.

\begin{figure}[!ht]
\centering
\includegraphics[scale=0.10]{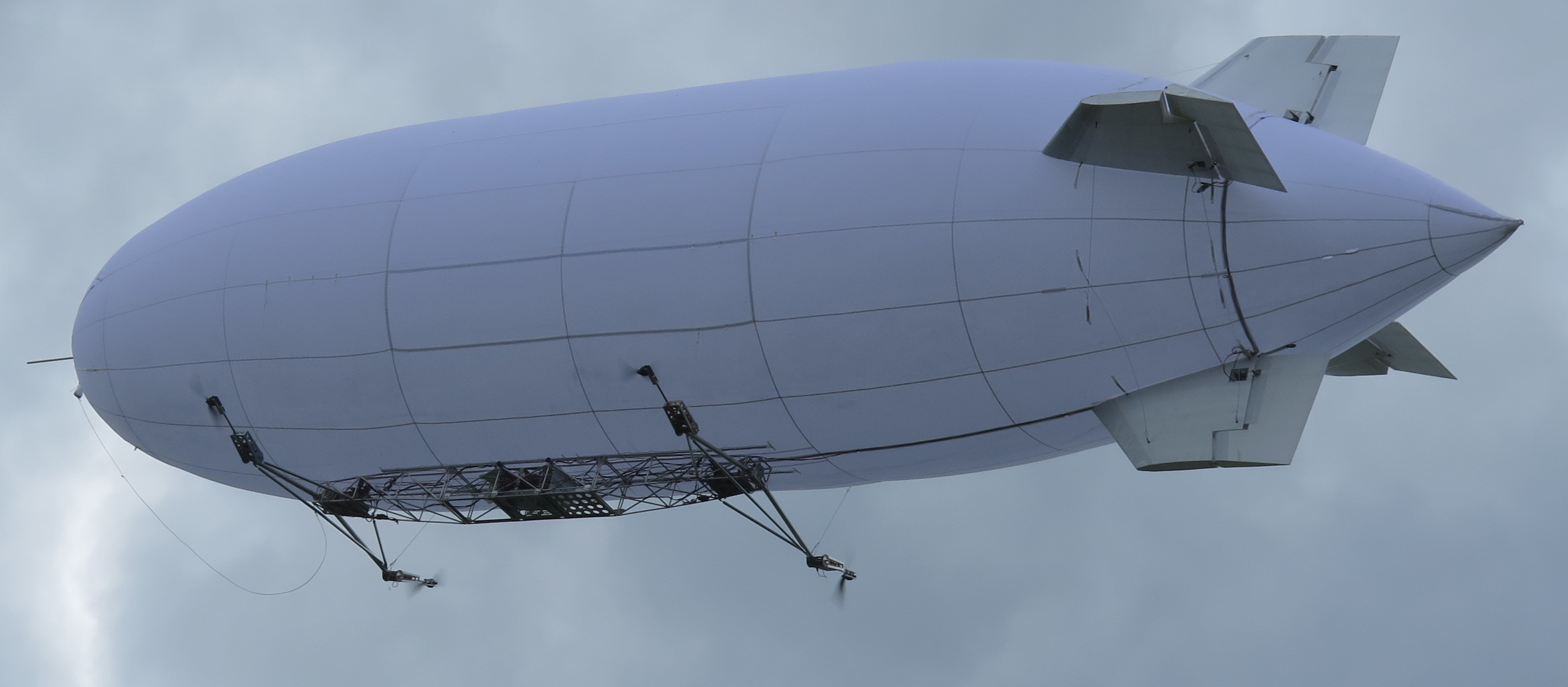}
\caption{NOAMAY airship for environmental monitoring in Amazon.\label{fig:principal:noamay}}
\end{figure}

For this kind of autonomy, a robust nonlinear control approach is necessary, in order to deal with the multiple and uncommon challenging specifications \cite{azinheira2009backstepping,moutinho2016airship}. This includes the highly nonlinear model and the flight-dependent dynamics, that ranges from the hovering to the aerodynamic flight, including VTOL, and the strong  wind/gust perturbations \cite{2012Bestaoui}.

In the last two decades, different nonlinear control methods have been proposed in the literature of airship control, where the most common are the sliding mode control (SMC) and the backstepping control (BS).
Sliding mode controllers \cite{1977Utkin,ES:98,2015SMCBook} are widely used due to their attractive features like  the insensitivity to matched uncertainties,
robustness to external perturbations, the simplicity and ease of implementation, and the finite-time convergence. 
And the attractive qualities of backstepping \cite{SL:91,FS:99} are the asymptotic global stability against parametric uncertainty, whether matched or mismatched, and the systematic recursive construction of  the Lyapunov functions.

In order to take advantage of the benefits of both control approaches, these techniques were combined or "mixed", in the beginning of the 1990s, to develop the "backstepping sliding mode controller" (BSMC) \cite{1997RiosZinober,1997Rios}. Indeed, still today, the scientific literature usually presents the BSMC approach as a simple union of the backstepping and the sliding mode techniques, without pointing out the drawbacks of putting together both approaches. Also, it is very hard to find a detailed analysis of the different possible BSMC variants regarding the choosing of the sliding surface,  the type of chattering smoothing, the presence/lack of cross-coupling cancellations in the backstepping procedure, the independence of the control gains adjust, the constraints on the gains to satisfy the Lyapunov function derivatives, etc. 
Further, it is interesting to note that most of the research reporting comparative analysis of BSMC x SMC or BSMC x BS, usually yield their conclusions based on a set of simulation/experimental results, without a rigorous analysis of the source of the findings, linking cause and effect. 






Regarding the literature on nonlinear control of airships, the pioneer works on backstepping appeared in France, in the scope of the LAAS/CNRS autonomous blimp project \cite{2002Hygounenc} and also in the LSC/Evry \cite{2002Beji,2012Bestaoui}.
But, as far as we know, the first work proposing a global approach  without spliting the control into longitudinal/lateral motions was the backstepping control (BS) of the AURORA airship \cite{azinheira2009backstepping}, considering both positioning  and path-tracking control for the whole flight envelope. However, the variables defined in the two steps of this backstepping do not follow the regular cascaded combination of position and velocity, avoiding the independence in gain adjusts \cite{FS:99}. Other works on backstepping control for airships can be found in \cite{2013Yang,2013Liesk,2017Liu}.
And regarding the sliding mode control for airships, it is also a very common solution found in the literature \cite{dPBB:07,2015YangANN,YWZ:16,moriguchi2017controle}.

For the BSMC project, it is important to say, firstly, that there are different kinds of approaches in the literature, regarding the type of sliding surface function considered, where the two most common are type 1, or BSMC-1 \cite{1997Rios} and type 2, or BSMC-2 \cite{1994Rios}. 
And for the airship  control using BSMC, we find few works in the literature.
In  \cite{YWZ:16}, the authors use a type-1 BSMC applied to a full actuated airship, comparing the results with a classical SMC controller.  However, as we show here, when using a type-1 BSMC, the independence of controller gains adjust is lost, which may affect the performance \cite{FS:99}. In a second work \cite{2017ABSMC}, another type-1 BSMC is used, with high adaptive gains, to assure tracking stability under bounded perturbations, but without control saturation constraints.
The third work \cite{2018Parsa} is a type-2 BSMC, like in our case, and the simulation results are used to demonstrate the improved performance of BSMC over single backstepping (BS), for rejecting wind disturbances. However, as the controller gains are not shown in their simulation results, further conclusions are not possible. And the fourth work \cite{2019BSMC} is a complex aproach,
where  adaptive sliding gains are designed to estimate the uncertainties, though not considering the actuator saturations.

In our work, we show that by using the type-2 BSMC, first proposed by Zinober in 1994 \cite{1994Rios}, and extending it to the MIMO case, it is possible to derive a common framework allowing for the design, in a same mathematical basis, of the three nonlinear control approaches (BS, SMC, BSMC), easing a theoretical-based comparative analysis. It is also important to say that we work with the classical BSMC that combines a first-order sliding mode (FOSMC) with a  regular backstepping, which is the most common and simple type of BSMC found in the literature, while other possible combinations, specially with high order SMC, are possible. Even so, many conclusions derived herein may be extended to some of the other cases.


Thus, the contribution of our paper is twofold. First, we present a new kind of vectorial backstepping sliding mode for the positioning and tracking of an autonomous airship, that maintains the advantages of both isolated techniques (BS and SMC). Second, we present a general framework basis for the design and analysis of vectorial BSMC-SMC-BS that can be applied to any dynamic system whose model is in “lower triangular block” form, like the mechanical systems with cascaded position-velocity variables.

After this introductory section, the remaining parts of this 
paper are organized as follows. Section II describes the airship modeling and the problem formulation. Section III presents the detailed  project for the vectorial BSMC, with the special features that allow to introduce the unified framework, shown in the end of the section. Section IV presents the comparative simulation results for the airship positioning and tracking. And Section V closes with conclusions and final remarks.

%% file: modeling.tex
\section{Airship Modeling and Problem Formulation}

An accurate mathematical model, together with a reliable simulator, are essential conditions for a successful control design. This section presents the full 6-DOF airship model used in the simulator software, as well as the kinematic  model and the problem formulation, with the proposed airship missions \cite{moutinho2016airship,azinheira2009backstepping}.

\vspace{-0.2cm}

\subsection{Airship Nonlinear Dynamic Model and Simulator}
\label{subsec:act}

\begin{figure}[!ht]
\centering
\includegraphics[scale=0.18]{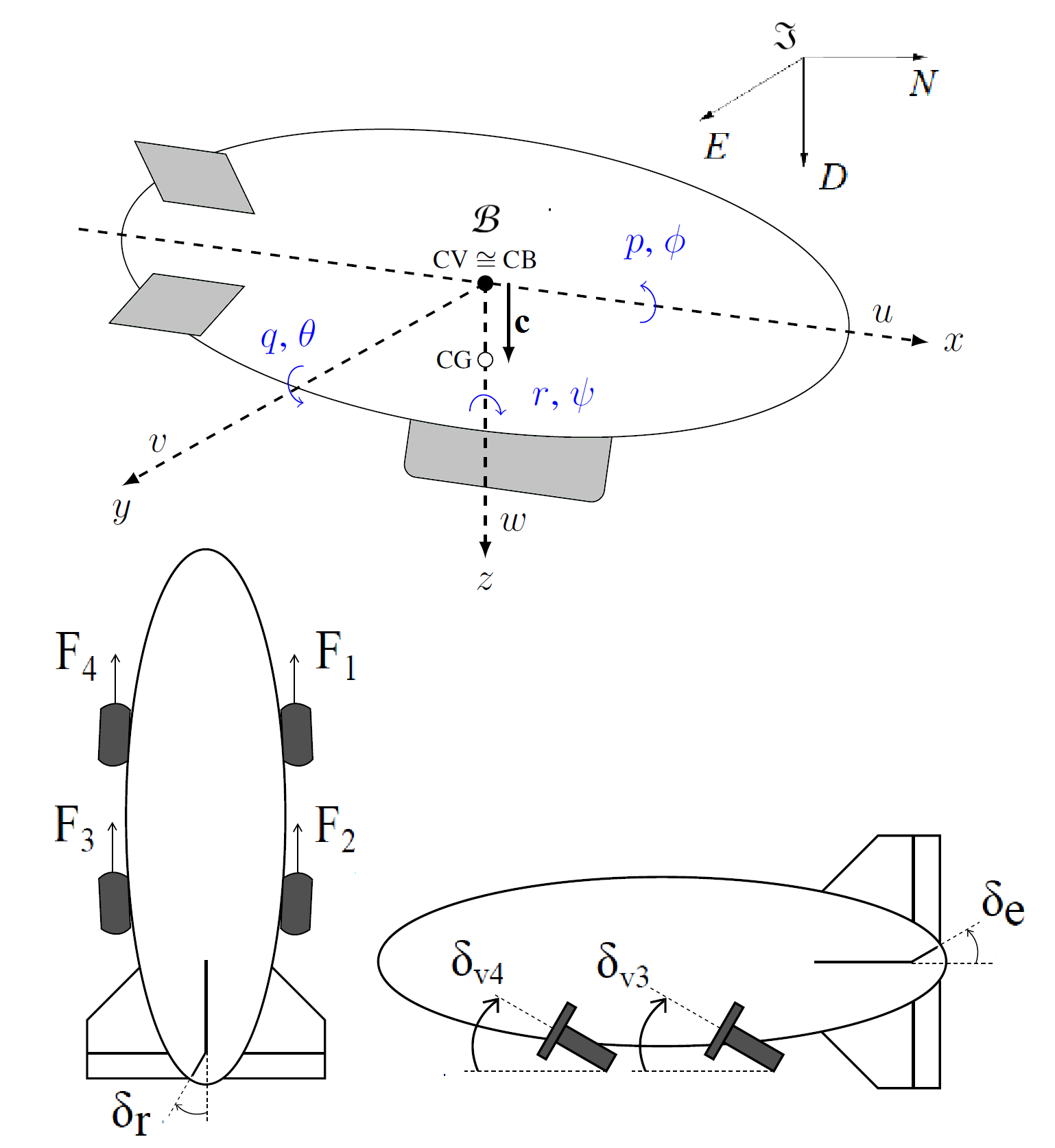}
\caption{Airship reference frames and actuators configuration.\label{fig:principal:ABC}}
\end{figure}

The Noamay airship (Figure \ref{fig:principal:noamay}) has $11m$ length, $2.55m$ in diameter, and $40m^3$ of volume. With an available payload of approximately  $6 kg$, its maximum speed is about  $55km/h$. As control actuators, the airship uses four vectorizable (tilting) electric thrusters and four tail surfaces (Figure \ref{fig:principal:ABC}).

It is assumed that the model considers the airship virtual masses/inertias due to the large volume of air displaced by the vehicle, and that the airship is a rigid body, so that aeroelastic effects are ignored. 

The dynamic model is defined in the local airship frame  $\mathcal{B}$. The orientation of this body-fixed frame $x, y, z$ with respect to an Earth-fixed frame $\mathfrak{I}$, with coordinates $P_N, P_E, P_D$, is obtained through the Euler angles, corresponding to the roll, pitch, and yaw angles, respectively. The airship linear and angular velocities are given by ${\bf{v}}= [u, v, w]^T$ and $\boldsymbol{\omega}=[p, q, r]^T$. The angular velocity components $p, q, r$ may also be considered as approximations of the roll, pitch, and yaw rates, respectively. The airship dynamics may then be expressed, from Newton's first law, as
\begin{equation}
\label{eq:mod:principal}
\mathbf{M}\dot{\mathbf{V}} = \mathbf{F}_k + \mathbf{F}_w + \mathbf{F}_g + \mathbf{F}_p + \mathbf{F}_a,
\end{equation}
where $ \mathbf{M} \in  \realnumber^{6\times6}$ is the generalized mass/inertia matrix; $ \matr{V} \in  \realnumber^{6} $ is composed of airship linear/angular velocities $ \matr{V} = [\matr{v}^T~~\boldsymbol{\omega}^T]^T $;
 $ \matr{F}_k \in  \realnumber^{6}  $ is the kinetics force with  centrifugal/Coriolis; $ \matr{F}_w \in  \realnumber^{6}  $ is the wind induced force; $ \matr{F}_g \in  \realnumber^{6}  $ is the gravity/buoyancy force; $ \matr{F}_p \in  \realnumber^{6}  $ is the propulsion force, and $ \matr{F}_a \in  \realnumber^{6}  $ is the aerodynamic force \cite{moutinho2016airship}.
 

 If we assume the wind as constant in the earth frame, the local velocity $\matr{v}$ may be expressed as the sum of the air velocity $\matr{v}_a$ (with modulus $V_t$) and the wind velocity in fixed frame $\matr{v}_w$, or $\matr{v} = \matr{v}_a + \matr{S}\matr{v}_w$, where ${\bf{S}}\in {R}^{3\times 3}$ represents the transformation matrix from inertial $\mathfrak{I}$ to body $\mathcal{B}$ frames. And 
defining the air velocity state vector as $ \matr{x} = [\matr{v}_a\tran~~\boldsymbol{\omega}\tran]\tran $, the airship dynamic equation in (\ref{eq:mod:principal}) may be expressed in a compact form as a function of this air velocity as \cite{AMdP:06,azinheira2009backstepping}:
\begin{equation}
\matr{M}\dot{\matr{x}} = -\boldsymbol{\Omega}_6\matr{M}\matr{x} + \matr{E}_g\matr{S}\matr{g} + \matr{F}_{a1} + \matr{f},
\label{eq:dinamica:airship:reformulada}
\end{equation}
where  $\matr{g}$ is the  gravity vector in the inertial frame, and $\matr{E}_{g}=\left[\begin{array}{c}
m_{w}\matr{I}_{3}\\
m\matr{C}_{3}
\end{array}\right]$,
with $m$ as the mass of the airship and $m_{w}$ as the weighting mass (excess of weight in relation to buoyancy).
Also,
$ \boldsymbol{\Omega}_6 \triangleq \mbox{diag}\{\boldsymbol{\Omega}_3,\boldsymbol{\Omega}_3\} \in \realnumber^{6\times6}$ 
is the synthetic matrix notation used to indicate cross-products, where $\matr{C}_{3}=\matr{c}\times \in \realnumber^{3\times3}$ and 
$\boldsymbol{\Omega}_{3}=\boldsymbol{\omega}\times \in \realnumber^{3\times3}$, with $\matr{c}\in\realnumber^{3}$  representing the CG coordinates in the local frame.

The force/moment vector $\matr{F}_{a1}$ in (\ref{eq:dinamica:airship:reformulada}) corresponds to the portion of the aerodynamic forces  $\matr{F}_{a}$ that is  dependent only on the aerodynamic angles, while $\matr {f}$ (control input) corresponds to the portion of $\matr{F}_{a}$ that depends also on the tail surface deflections as well as the propulsion force $\matr{F}_p$.

When using the control allocation, the force input can be transformed into the actuators input vector 
$	{u} = [{\delta}_e \ {\delta}_r \ {\delta}_1 \ {\delta}_2 \ {\delta}_3 \ {\delta}_4 \ {\delta}_{v1} \ {\delta}_{v2} \ {\delta}_{v3} \ {\delta}_{v4}]^T
$ which includes the tail aerodynamic surface deflections (${\delta}_e \ {\delta}_r$), the engine thrust inputs (${\delta}_i$) and the engines vectoring angles (${\delta}_{vi}$) \cite{vieira2017controle}.


Note  also that the four propellers (Figure \ref{fig:principal:ABC}) can work in four different configurations, regarding the differential thrust used, from the available independent motors actuation.

The airship kinematic model is represented as follows \cite{moutinho2016airship}.
The global position vector of the airship in space is defined by $ \boldsymbol{\eta} = [\matr{p}\tran ~~\matr{q}\tran]\tran \in \realnumber^{7}$, where $ \matr{p} =[P_N, P_E, P_D]^T$ represents the coordinates in the inertial frame $\mathfrak{I}$,  and $ \matr{q} \in \realnumber^4$ is the  attitude (with quaternions), whose derivatives are given by:

\begin{equation}
 \begin{array}{l}
	\dot{\matr{p}}=\matr{S}\tran \matr{v}~~~~~~~~~~~~~~
	\dot{\matr{q}}=\frac{1}{2}\matr{Q}\left[\begin{array}{c}
	0\\
	\boldsymbol{\omega}
	\end{array}\right]
		\end{array},
	\label{eq:7D-kinematics-1}
\end{equation}
where $\matr{Q}\in\realnumber^{4\times4}$ is the unitary matrix that relates quaternions to their derivatives and angular rates, and ${\bf{S}}\in \textsf{SO}(3)=\{{\bf{S}}\in {R}^{3\times 3} | {\bf{S}}^T{\bf{S}}={\bf{I}},\det({\bf{S}})=1\}$ is the transformation matrix converting linear velocities from inertial to local frames, whose elements are functions of the quaternions and $\matr{Q}$ matrix \cite{azinheira2006airship}. 

Such that the complete kino-dynamic equations of the system can be finally expressed by:
\begin{equation}
\begin{array}{l}
\dot{\matr{x}} =\matr{K}\matr{x}+\matr{M}^{-1}\left(\matr{E}_{g}\matr{S}\matr{g}+\matr{F}_{a1}+\matr{f}\right),\\
\dot{\boldsymbol{\eta}} =\matr{T}\matr{x}+\matr{B}\matr{v}_{w},\\
\ddot{\boldsymbol{\eta}} = \matr{T}\dot{\matr{x}}+\matr{D}\boldsymbol{\Omega}_7\matr{C}\matr{x},
\end{array}
\label{eq:final:aceleracao}
\end{equation}
where
 $\matr{C}_{7\times6}$ is a constant matrix used to match the different dimensions in the matricial equation, and  $\matr{D}_{7\times7}$ is a matrix containing  ${\bf{S}}^T$ and  $\frac{1}{2}\matr{Q}$ in its diagonal. Also,
$ \matr{K} \triangleq -\matr{M}^{-1}\boldsymbol{\Omega}_6\matr{M} \in \realnumber^{6\times6} $ is linearly dependent on the angular velocity 
$ \boldsymbol{\omega} $, and  the mass matrix $\matr{M} $ is assumed constant or slowly varying with altitude (since buoyancy depends on the air density), and  
$\boldsymbol{\Omega}_7 = \mbox{diag}(\boldsymbol{\Omega}_3,\frac{1}{2}\boldsymbol{\Omega}_4)$, as shown in \cite{moutinho2016airship,azinheira2009backstepping}.

\subsection{Problem Formulation}
\label{sec:typemissions}
The objective is to derive three nonlinear control laws (BS,SMC,BSMC) in a single unified framework formulation, to assure stability and performance for the autonomous airship control in the following three tasks: positioning, path tracking and velocity tracking.
Depending on the specific task or flight mission, the position/velocity errors should be specified accordingly, with appropriate references.

With this purpose, let us define $ \boldsymbol{\eta}_d = [\matr{p}\tran_d ~~\matr{q}\tran_d]\tran \in \realnumber^{7}$ as the airship inertial reference position, and $ \matr{x}_d = [\matr{v}\tran_d ~~ \matr{0}]\tran \in \realnumber^{6}$ as the airship ground velocitiy reference, as well as $ \tilde{\boldsymbol{\eta}} = \boldsymbol{\eta} - \boldsymbol{\eta}_d \in \realnumber^{7}$ as the position error. Then, the definition of the errors for the controllers in each type of mission is:
\begin{enumerate}
    \item[1.] Positioning - Here $ \boldsymbol{\eta}_d=\mbox{constant}$, and therefore  $ \dot{\tilde{\boldsymbol{\eta}}} = \dot{\boldsymbol{\eta}} $.
    \item[2.] Path tracking - In this case, the airship reference 
    would contain the trajectory profile, as a function of time, or $ {\boldsymbol{\eta}}_d(t) $, such that  $ \tilde{\boldsymbol{\eta}} = {\boldsymbol{\eta}} - {\boldsymbol{\eta}}_d$ and $ \dot{\tilde{\boldsymbol{\eta}}} = \dot{\boldsymbol{\eta}} - \dot{\boldsymbol{\eta}}_d  $.
    \item[3.]  Velocity tracking - Here, a constant ground velocity $ \matr{v}_d $ is to be tracked in a rectilinear path. In this case we have $ \tilde{\boldsymbol{\eta}} = {\boldsymbol{\eta}} - {\boldsymbol{\eta}}_d$ and $ \dot{\tilde{\boldsymbol{\eta}}} = \dot{\boldsymbol{\eta}} - \dot{\boldsymbol{\eta}}_d  $, in specific $ \dot{\boldsymbol{\eta}}_d = \begin{bmatrix}
	\matr{S}_d\matr{v}_d & \matr{0}
	\end{bmatrix}^{\footnotesize\mbox{T}} = \matr{T}_d\matr{x}_d$, where $\matr{T}_d = \begin{bmatrix}
	\matr{S}_d & \matr{0}
	\end{bmatrix}^{\footnotesize\mbox{T}}$.
\end{enumerate}


With respect to the NOAMAY airship simulator (Simulink/Matlab), it is an evolution of the previous AURORA   model/simulator \cite{depaiva2006,moutinho2016airship}, with the inclusion of the new propulsion model of 4-electrical tilting thrusters \cite{vieira2017controle}.

%% file: controls.tex
\section{Nonlinear Control Framework Design }

In this section, we present the proposed block backstepping-sliding mode approach (BSMC), that is the basis for the unified framework control design.




\input{part/controls/bsmc.tex}

\input{part/controls/BSMC2.tex}
\input{part/controls/unified.tex}

%% file: part/controls/bsmc.tex
\subsection{Backstepping Sliding Mode Control Preliminaries}
\label{subsec:bsmc}	

As for the regular backstepping (BS) or sliding mode (SMC) projects, the BSMC technique can also appear in the SISO and MIMO versions, where the last one is known as ``vectorial BSMC'' or ``{block backstepping sliding modes}'' \cite{YWZ:16}. The block backstepping technique has emerged as one of the most efficient algorithms based on backstepping \cite{FS:99,2017Liu}, keeping the advantage of the selective nonlinear cancellation, but without the strict feedback constraint in the state variables. Indeed, the  block backstepping only  requires that the dynamic matrix  be in ``lower triangular block'' form \cite{FS:99}.

For didatic purposes, we will start showing some preliminaries on BSMC for the SISO case, to better illustrate the influence of the choosing of the sliding surface, as well as some fundamental properties used in the unified design. 

Thus, without loss of generality, we will consider, in this first case, the SISO nonlinear cascaded model of:
\begin{equation}
\left.\left\{ \begin{array}{l}
\dot{x}_{i}=x_{i+1},\,\,\,i=1,...,n-1\\
\dot{x}_{n}=f_{n}(x_{1},x_{2},\ldots,x_{n})+g_{n}(x_{1},x_{2},\ldots,x_{n})u\\
y=x_{1}
\end{array}\right.\right.\label{eq:eq: PFModel-1-1b}
\end{equation}

Also, we summarize below the two most common first-order classical BSMC projects found in the literature, that will be denoted here by BSMC-1 \cite{1997Rios,1997RiosZinober} and BSMC-2 \cite{1994Rios}, where this second one is the one adopted in our work. 
\begin{singlespace}
\[
\begin{array}{ll}
& {\mbox{\textbf{BSMC-1 Project  (Rios-Bolivar~~et~~ al.,~~ 1997)}}} ~\\ \\
{\mbox{(1)}} &
{\textcolor{red}{\sigma=c_{1}z_{1}+c_{2}z_{2}+\ldots+c_{n-1}z_{n-1}+z_{n}}}\\
{\mbox{(2)}}&
\dot{\sigma}=c_{1}\dot{z}_{1}+c_{2}\dot{z}_{2}+\ldots+c_{n-1}\dot{z}_{n-1}+f_{n}+g_{n}u-\dot{\alpha}_{n-1}\\
{\mbox{(3)}}  &
u=\frac{1}{g_{n}}[-f_{n}-c_{1}\dot{z}_{1}-c_{2}\dot{z}_{2}-\ldots-c_{n-1}\dot{z}_{n-1}+ \\
& +\dot{\alpha}_{n-1}-\lambda_{n}\sigma-\rho\sgn(\sigma)\\
{\mbox{(4)}}
& \sigma\dot{\sigma}=\sigma(-\lambda_{n}\sigma-\rho\sgn(\sigma))=-\lambda_{n}\sigma^{2}-\rho\left|\sigma\right|\\
{\mbox{(5)}}  & V_{n}=V_{n-1}+\frac{1}{2}\sigma^{2}=\frac{1}{2}\sum_{i=1}^{n-1}z_{i}^{2}+\frac{1}{2}\sigma^{2}\\
{\mbox{(6)}}  &
\dot{V}_{n}=\dot{V}_{n-1}+\sigma\dot{\sigma}=-\sum_{i=1}^{n-1}k_{i}z_{i}^{2}+ {z_{n-1}z_{n}}+\sigma\dot{\sigma}\\
 &
\dot{V}_{n}=-\sum_{i=1}^{n-1}k_{i}z_{i}^{2} {+z_{n-1}z_{n}}-\lambda_{n}\sigma^{2}-\rho\left|\sigma\right|
\end{array}
\]

\[
\begin{array}{ll}
& {\mbox{\textbf{BSMC-2 Project (Zinober \& Rios-Bolivar, 1994)}}} \\ \\
{\mbox{(1)}} &
{\textcolor{red}{\sigma=z_{n}}}\\
{\mbox{(2)}}&
\dot{\sigma}=f_{n}+g_{n}u-\dot{\alpha}_{n-1}\\
{\mbox{(3)}}  &
u=\frac{1}{g_{n}}\left[-f_{n}{{-\lambda_{n-1} z_{n-1}}}+\dot{\alpha}_{n-1}-\lambda_{n}\sigma-\rho\sgn(\sigma)\right]\\
{\mbox{(4)}} & \sigma\dot{\sigma}~~=~~\sigma({-\lambda_{n-1} z_{n-1}}~-\lambda_{n}\sigma~-\rho\sgn(\sigma))=\\
& =~{-\lambda_{n-1} z_{n-1} z_{n}}~-\lambda_{n}z_{n}^{2}~-\rho\left|{z_n}\right| \\
{\mbox{(5)}}  & V_{n}=V_{n-1}+\frac{1}{2}\sigma^{2}=\frac{1}{2}\sum_{i=1}^{n-1}z_{i}^{2}+\frac{1}{2}\sigma^{2}\\
{\mbox{(6)}}  &
\dot{V}_{n}=\dot{V}_{n-1}+\sigma\dot{\sigma}=-\sum_{i=1}^{n-1}k_{i}z_{i}^{2}+z_{n}\dot{z}_n\\
 &
\dot{V}_{n}=-\sum_{i=1}^{n-2}k_{i}z_{i}^{2}-\lambda_{n-1}k_{n-1}z_{n-1}^2-\lambda_{n}z_{n}^{2}-{\rho}\left|z_n\right|
\end{array}
\]
\end{singlespace}
\noindent
\textbf{(1) Sliding function; (2) Sliding function derivative; (3) Control output; 
(4) $\sigma\dot{\sigma}$ (5) Lyapunov function; (6) Lyapunov function derivative.}
~\\

Where $z_{i}$ are the error variables from the 
backstepping procedure, and the design parameters
$c_{i}>0,\,i=1,\ldots,n-1$, are chosen such that  $p(s)=c_{1}+c_{2}s+\ldots+c_{n-1}s^{n-2}+s^{n-1}$, in the complex variable $s$, is Hurwitz. Moreover, $k_{i},\,i=1,\ldots,n$ are positive design gains, as well  $\rho$ and $\lambda$  \cite{1997RiosZinober}.

The main source of difference between BSMC-1 and BSMC-2 is the definition of the sliding surface manifold (in red, above). While for BSMC-1 we have $\sigma_{B1}$ as a linear combination of the errors, for BSMC-2 we have  $\sigma_{B2}$ defined simply as the last error $z_n$.

Prior to make a comparative discussion on both types of BSMC, we recall first two classical results, one in backstepping control \cite{FS:99}, and the other on regular sliding mode control \cite{2015SMCBook}, as follows.
~\\ \\
\noindent
{\textbf{Property 1 - Backstepping cross-coupling cancellation}} (Fossen, 1999)
~\\ 
In the regular backstepping approach (as well as in the block backstepping), the cross-coupling
cancellation of the error terms in the derivative of the Lyapunov function $V_{n}=\frac{1}{2}\sum_{i=1}^{n}z_{i}^{2}$ occurs due to the recursive procedure of construction of the error variables \cite{FS:99}, that is:
\begin{equation}
\begin{array}{l}
\dot{V}_{n}=z_{1}\dot{z}_{1}+z_{2}\dot{z}_{2}+\ldots+z_{n-1}\dot{z}_{n-1}+z_{n}\dot{z}_{n}\\
\dot{V}_{n}=z_{1}(-k_{1}z_{1}+z_{2})+z_{2}(-{z_{1}}-k_{2}z_{2}+z_{3})+\ldots+\\
\,\,\,+\ldots+z_{n}(-z_{n-1}-k_{n}z_{n})\\
\dot{V}_{n}=-(k_{1}z_{1}^{2}+k_{2}z_{2}^{2}+\ldots+k_{n}z_{n}^{2})<0.
\end{array}
\label{eq:prop1}
\end{equation}
~\\ 
 {\bf{Property 2 - Sliding Mode Atraction and Invariance}} (Shtessel et al., 2015) ~\\ 
In the standard first order classical sliding mode control \cite{2015SMCBook}, the condition for the state to reach the sliding surface $\sigma$ in finite time $t_{r}$ (atraction), and remain on it after that (invariance) is:
\begin{equation}
\sigma\dot{\sigma}<0\,\,\,\textrm{for\,\,}\:t<t_{r}\,\textrm{\,\,\,and\,\,\, }\sigma=\dot{\sigma}=0\textrm{ \,\,for\,\,\,}t\geq t_{r}.
\label{eq:prop2}
\end{equation}

~ \\
\indent
Thus, the different surface definitions for BSMC1 and BSMC2 imply different conditions for the reaching/invariance property of the sliding surfaces (Property 2). While for BSMC-1 we have $\sigma_{B1}\dot{\sigma}_{B1}=-\lambda_{n}\sigma^{2}-\rho\left|\sigma\right|<0$, for BSMC-2 we have $\sigma_{B2}\dot{\sigma}_{B2}={-\lambda_{n-1} z_{n}z_{n-1}}-\lambda_{n}\sigma^{2}-\rho\left|\sigma\right|$ which may not be negative, due to $z_{n}z_{n-1}.$
Such that attraction/invariance sliding property for BSMC-2  will hold only if we use a special time varying switching gain $\rho(t)$, as shown in the sequel. Otherwise, if a constant $\rho$ is used, the BSMC-2 project will exhibit a kind of ``dual behavior''. In this case, the state variable errors
$z_{i}$ will be attracted by the origin of the state space, like in a regular backstepping, for large initial state errors, and will be attracted by the sliding manifold, like in a regular SMC, when the state errors diminish to a given value.

Indeed, the critical switching gain for the transition between these two behaviors is  given by:
$\rho_{crit}=\lambda_{n-1}\left|z_{n-1}\right|$ , as if $\rho>\lambda_{n-1}\left|z_{n-1}\right|$ we have
\[
\begin{array}{l}
\sigma_{B2}\dot{\sigma}_{B2}=-\lambda_{n-1} z_{n-1}z_{n}-\lambda_{n}\sigma^{2}-\rho\left|z_{n}\right|\leq \\
\leq  \rho_{crit}\left|z_{n}\right|-\lambda_{n}\sigma^{2}-\rho\left|z_{n}\right|=-\lambda_{n}\sigma^{2}-(\rho-\rho_{crit})\left|z_{n}\right|<0
\end{array}
\]

Thus, if we use a time-varying switching gain like $\rho(t)=\rho_{crit}(t)+{\rho}_0=\lambda_{n-1}\left|z_{n-1}\right|+{\rho}_0$, then we always have $\rho>\rho_{crit}$ with $\sigma_{B2}\dot{\sigma}_{B2}<0$, and the reaching property with invariance will hold for BSMC-2. 

Another important point here is that the sliding surface of BSMC-2 project ($\sigma_{B2}=z_{n})$ is a very simple relation \cite{1994Rios} when compared to the one of BSMC-1 ($\sigma_{B1}=c_{1}z_{1}+\ldots+c_{n-1}z_{n-1}+z_{n}$). At a first glance, it could seem that this simple function would not capture all the information from the system error variables. However, due to
the recursive construction form of the error variables in the backstepping
procedure $(\dot{z}_{i}=z_{i+1}-k_{i}z_{i}-z_{i-1})$, the sliding
manifold function for BSMC-2 is indeed a linear combination of the primary
error $z_{1}$ and its $n-1$ derivatives. For example, 
considering the particular SISO cases of $n=2$, and $n=3$, we have:
\[
\begin{array}{ll}
n=2 & \sigma_{B2}=z_{n}=z_{2}=\dot{z}_{1}+k_{1}z_{1}\\
n=3  & \sigma_{B2}=z_{n}=z_{3}=\ddot{z}_{1}+(k_{1}+k_{2})\dot{z}_{1}+({\lambda}+k_{1}k_{2})z_{1}
\end{array}
\]

Such that the essential difference in the sliding surfaces definitions 
is that for BSMC-1 the designer has additional degrees of freedom in
the manifold parameters design, which is not indeed a big drawback for BSMC-2.
Care should be taken, however, in the selection of the controller gains $k_{i}$, in order to assure that the final polynomial derived from the linear combination of $z_1$ and its derivatives is Hurwitz.

And, as the cross-coupling cancellation  of all the $z_{i}$ terms (Property 1) occurs for BSMC-2, we still keep the independence of the adjustment of the controller gains like in the pure backstepping case. Note that this is not the case for BSMC-1, when the cross-coupling cancellation does not hold for the last step. The consequence is that the resulting control output for BSMC-1 does not show the term $\lambda_{n-1} z_{n-1}$, which is responsible for the gain independence and the performance enhancement, as we will see in the next section.

%% file: part/controls/BSMC2.tex
\subsection{MIMO Backstepping Sliding Mode Design}
\label{subsec:bsmc2}	

Now, we can return to the MIMO design procedure for the {block backstepping sliding modes} which follows the same rules of a regular backstepping, except for the last step, when a new variable is introduced, related to the sliding mode feature.

First, let us define the two block-state variables $\matr{x}_{1}=\boldsymbol{\eta}$ and $\matr{x}_{2}=\dot{\boldsymbol{\eta}}$,
following the previous airship model equation (\ref{eq:final:aceleracao}), such that we have:
\begin{equation}
\left\{ \begin{array}{l}
{\dot{\matr{x}}_{1}}  =\mathbf{x_{2}}\\
{\dot{\matr{x}}_{2}}  =\mathbf{T\dot{x}+D\varOmega_{7}Cx}=\\
 =\underbrace{\matr{T}~[\matr{K}\matr{x}+\matr{M}^{-1}\left(\matr{E}_{g}\matr{S}\matr{g}+\matr{F}_{a1}\right)]+\matr{D}\boldsymbol{\Omega}_7\matr{C}\matr{x}}_{\matr{f}_{2}(\matr{x}_{1},\matr{x}_{2})}+\underbrace{\matr{T}\matr{M}^{-1}}_{\matr{g}_{2}(\matr{x}_{1},\matr{x}_{2})}\matr{u},
\end{array}\right.
\label{eq:modx1}
\end{equation}
where $\matr{u}=\matr{f}$ is the control vector (forces/moments).

The position/velocity tracking errors for the airship positioning/path tracking tasks are defined as:
\begin{equation}
\begin{array}{l}
\tilde{\boldsymbol{\eta}} = \boldsymbol{\eta} - \boldsymbol{\eta}_d,\\
\dot{\tilde{\boldsymbol{\eta}}} = \dot{\boldsymbol{\eta}} - \dot{\boldsymbol{\eta}}_d,
\end{array}
\label{eq:imple:1}
\end{equation}
where $ \boldsymbol{\eta}_d $ is the desired trajectory profile. 

\begin{figure}[!ht]
\centering
\includegraphics[scale=0.42]{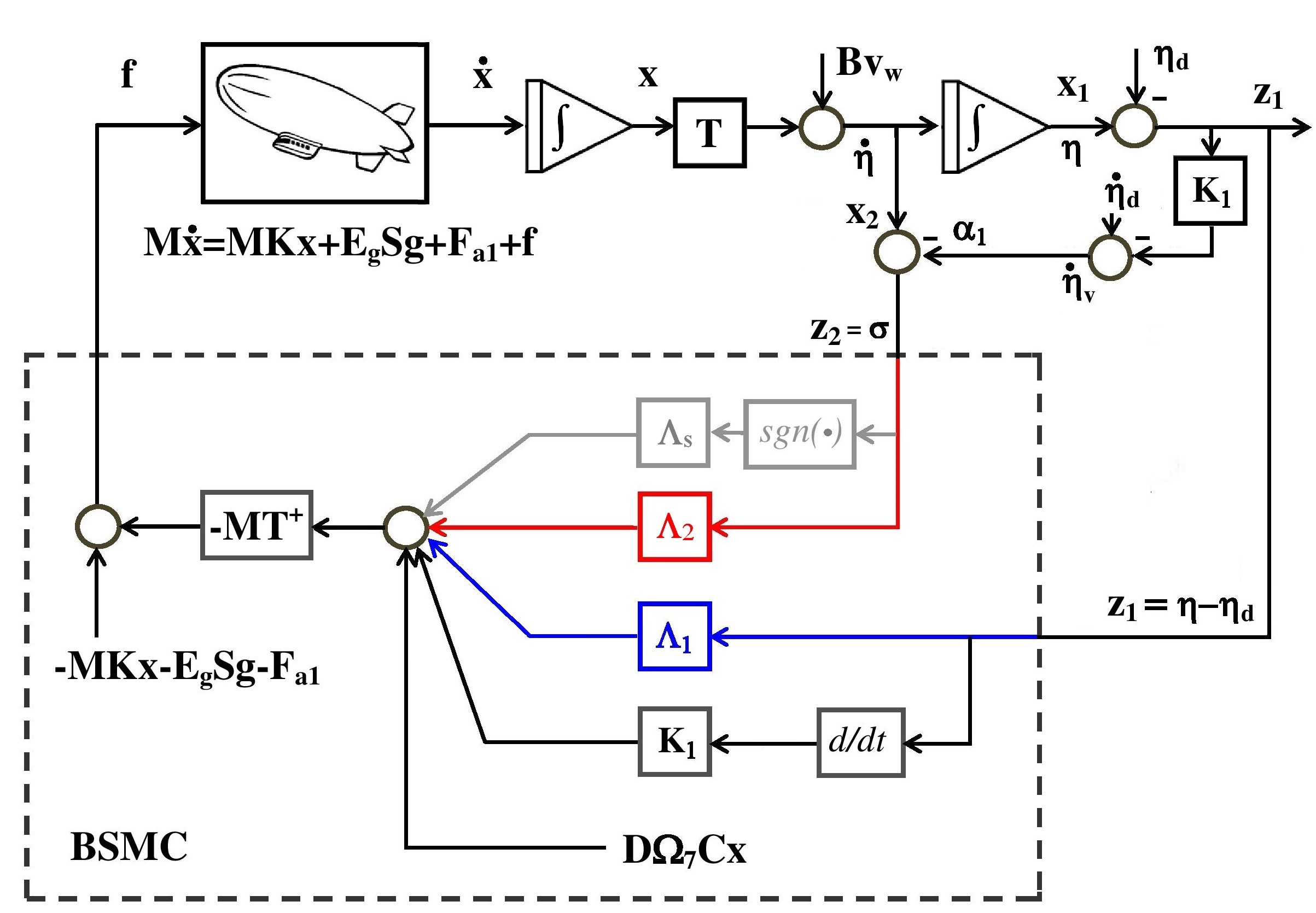}
\caption{Block diagram for closed-loop system with BSMC controller.\label{fig:principal:blocks}}
\end{figure}

Moreover, we define the so called "virtual trajectory" reference velocity ${\dot{\boldsymbol{\eta}}}_v$, as proposed in \cite{SL:91}, such that we have:
\begin{equation}
\begin{array}{l}
\dot{\boldsymbol{\eta}}_v = \dot{\boldsymbol{\eta}}_d - \matr{K}_1\tilde{\boldsymbol{\eta}}
\end{array},
\label{eq:imple:3}
\end{equation}
where $ \matr{K}_1 $ is a positive definite gain matrix with compatible units. Look that when the position error is null ($\tilde{\boldsymbol{\eta}}=0$), the virtual trajectory velocity ${\boldsymbol{\eta}}_v$ will be equal to the desired velocity ${\boldsymbol{\eta}}_d$ (see Fig. \ref{fig:principal:blocks}). This adaptation, as proposed in \cite{SL:91}, is necessary to transform a system with second order relative degree into one of first order relative degree.

Such that the two steps of this two-block BSMC design can be derived in the following way.

\noindent
\textbf{STEP 1.}

If we define the first backstepping variable as the position error, or
\begin{equation}
	\matr{z}_1 = \tilde{\matr{x}}_1 = \tilde{\boldsymbol{\eta}},
	\label{eq:new:variavel:bs}
\end{equation}
then, the  vector state $ \matr{x}_2 $ will be the virtual control input defined by:
\begin{equation}
\begin{array}{l}
\matr{x}_2 \triangleq \matr{z}_2 + \boldsymbol{\alpha}_1,
\end{array}
\label{eq:imple:2}
\end{equation}
where $ \matr{z}_2 $ is the new transformed vector error, and $ \boldsymbol{\alpha}_1 $ is the stabilization vector, that will be made equal to the "virtual trajectory" reference velocity ${\dot{\boldsymbol{\eta}}}_v$, or:
\begin{equation}
\begin{array}{l}
\boldsymbol{\alpha}_1 = 
\dot{\boldsymbol{\eta}}_v = \dot{\boldsymbol{\eta}}_d - \matr{K}_1\tilde{\boldsymbol{\eta}}
\end{array},
\label{eq:imple:3b}
\end{equation}
Thus, following the previous equations, the new backstepping variable $ \matr{z}_2 $, and its time derivative, are:
\begin{equation}
\begin{array}{l}
\matr{z}_2  = {\matr{x}}_2 - {\boldsymbol{\alpha}}_1 =
\dot{{\boldsymbol{\eta}}}-( \dot{\boldsymbol{\eta}}_d - \matr{K}_1{\tilde{\boldsymbol{\eta}}})=
\dot{\tilde{\boldsymbol{\eta}}} + \matr{K}_1\tilde{\boldsymbol{\eta}}= \dot{\matr{z}}_1+\matr{K}_1{\matr{z}}_1 \\
\dot{\matr{z}}_2=\dot{\matr{x}}_2 -\dot{\boldsymbol{\alpha}}_1=
\matr{f}_{2}+\matr{g}_{2}\matr{u}-\dot{\boldsymbol{\alpha}}_1
\end{array}
\label{eq:imple:4}
\end{equation}


Now, still in this first step of the backstepping design, we define a candidate Lyapunov function and its derivative as:
\begin{equation}
\begin{array}{c}
{v}_1 = \frac{1}{2}\matr{z}_1^{\footnotesize\mbox{T}}\boldsymbol{\Lambda}_1\matr{z}_1,\\
\dot{{v}}_1 = \matr{z}_1^{\footnotesize\mbox{T}}\boldsymbol{\Lambda}_1{\color{black}\dot{\matr{z}}_1}, \\
\end{array}
\label{eq:imple:5}
\end{equation}
where $ \boldsymbol{\Lambda}_1 $ is a positive definite diagonal matrix ($ \boldsymbol{\Lambda}_1 = \boldsymbol{\Lambda}_1^{\footnotesize\mbox{T}}  $).

And deriving $\matr{z}_1$ from ~\eqref{eq:new:variavel:bs}, using  \eqref{eq:imple:4} and  \eqref{eq:imple:5} we have:
\begin{equation}
\begin{array}{c}
\dot{{v}}_1 = \matr{z}_1^{\footnotesize\mbox{T}}\boldsymbol{\Lambda}_1\dot{\matr{z}}_1=\matr{z}_1^{\footnotesize\mbox{T}}\boldsymbol{\Lambda}_1{(\matr{z}_2 - \matr{K}_1{\matr{z}}_1)}=-\matr{z}_1^{\footnotesize\mbox{T}}\boldsymbol{\Lambda}_1\matr{K}_1\matr{z}_1 + \matr{z}_1^{\footnotesize\mbox{T}}\boldsymbol{\Lambda}_1\matr{z}_2
\end{array}
\label{eq:imple:7}
\end{equation}
and, consequently, as the first parcel is negative, the first system (with $ \matr{z}_1$ error) will be stabilized, provided that the second error converges ($ \matr{z}_2 = \matr{0} $).

\textbf{STEP 2}. Set the sliding surface as $\boldsymbol{\sigma}=\matr{z}_{2}$.

Now, if we recall that $\matr{x}_2={\matr{z}}_{2}+\boldsymbol{\alpha}_{1}$, from  \eqref{eq:imple:2}, and
$\dot{\matr{x}}_2=\matr{f}_{2}+\matr{g}_{2}\matr{u}$, from \eqref{eq:modx1}, we have that the derivative of the sliding function is given by: 
$\dot{\boldsymbol{\sigma}}=\dot{\matr{z}}_{2}=\matr{f}_{2}+\matr{g}_{2}\matr{u}-\dot{\boldsymbol{\alpha}}_{1}$. Thus we have the associated Lyapunov function and its derivative as: 
\begin{equation}
\begin{array}{c}
v_{2}= \frac{1}{2}\matr{z}_1^{\footnotesize\mbox{T}}\boldsymbol{\Lambda}_1\matr{z}_1+
\frac{1}{2}\matr{z}_{2}^{\footnotesize\mbox{T}}\matr{z}_{2} \\
\dot{v}_{2}=\matr{z}_{1}^{\footnotesize\mbox{T}}\boldsymbol{\Lambda}_1\dot{\matr{z}}_{1}+\matr{z}_{2}^{\footnotesize\mbox{T}}\dot{\matr{z}}_{2}
=\matr{z}_{1}^{\footnotesize\mbox{T}}\boldsymbol{\Lambda}_1\dot{\matr{z}}_{1}+\matr{z}_{2}^{\footnotesize\mbox{T}}
[\matr{f}_{2}+\matr{g}_{2}\matr{u}-\dot{\boldsymbol{\alpha}}_{1}] \\
     \dot{v}_{2}=-\matr{z}_1^{\footnotesize\mbox{T}}\boldsymbol{\Lambda}_1\matr{K}_1\matr{z}_1 + \matr{z}_1^{\footnotesize\mbox{T}}\boldsymbol{\Lambda}_1\matr{z}_2
    {-\matr{z}_{1}^{\footnotesize\mbox{T}}}{\boldsymbol{\Lambda}_{1}\matr{z}_{2}}-{\matr{z}_{2}^{\footnotesize\mbox{T}}}{\boldsymbol{\Lambda}_{2}\matr{z}_{2}}-\rho\Vert  \matr{z}_{2}\Vert_{1}
    \end{array}
\label{eq:LyapGeral}
\end{equation}


Where we have applied the control law given by 
${\matr{u}=\matr{g}_{2}^{-1}[-\matr{f}_{2}-{\boldsymbol{\Lambda}_{1}}\matr{z}_{1}}+\dot{\boldsymbol{\alpha}}_{1}-{\boldsymbol{\Lambda}_{2}\matr{z}_{2}}-\rho\sgn({\matr{z}_{2}})],$
with $\boldsymbol{\Lambda}_{2}$ positive definite diagonal matrix, and the scalar $\rho >0$.
Thus, the Lyapunov derivative $\dot{v}_{2}$ is negative, and in this way, the regulation at the origin is ensured since $\lim_{t\rightarrow\infty}{\matr{z}_{i}}=0,~i=1,~2,\ldots,~n$, and in particular $\lim_{t\rightarrow\infty}{\matr{z}_{1}}=0.$ 

However, as in the SISO case, note that the sliding invariance property of the block-BSMC is not guaranteed, as
\[
\boldsymbol{\sigma}^{\footnotesize\mbox{T}}\dot{\boldsymbol{\sigma}}=
-{\matr{z}_{2}^{\footnotesize\mbox{T}}} \dot{\matr{z}}_{2}=
{-\matr{z}_{2}^{\footnotesize\mbox{T}}}{\boldsymbol{\Lambda}_{1}\matr{z}_{1}}-{\matr{z}_{2}^{\footnotesize\mbox{T}}}{\boldsymbol{\Lambda}_{2}\matr{z}_{2}}-\rho\Vert  \matr{z}_{2}\Vert_{1}
\]
may not be negative.

Thus, as before, if one wants to assure the invariance property for this BSMC design, it is necessary to define a time-varying switching gain $\rho(\matr{z})$. For this purpose, consider:
\[
\boldsymbol{\sigma}^{\footnotesize\mbox{T}}\dot{\boldsymbol{\sigma}}=-{\boldsymbol{\sigma}^{\footnotesize\mbox{T}}{\boldsymbol{\Lambda}}_{2}\boldsymbol{\sigma}-\boldsymbol{\sigma}^{\footnotesize\mbox{T}}({\boldsymbol{\Lambda}}_{1}\matr{z}_{1}}+\rho\sgn({\boldsymbol{\sigma}})),
\]
and as ${\boldsymbol{\Lambda}}_{1}$ is a diagonal matrix, the second term in the right side of the above equation can be written as a function of the individual elements of the respective vectors, or:
\[
\boldsymbol{\sigma}^{\footnotesize\mbox{T}}\dot{\boldsymbol{\sigma}}=-{\boldsymbol{\sigma}^{\footnotesize\mbox{T}}{\boldsymbol{\Lambda}}_{2}\boldsymbol{\sigma}-}\left[\begin{array}{c}
\sigma_{i}({\Lambda_{1}^{i,i}}z_{1}^{i}+\rho\sgn(\sigma_{i}))\\
\end{array}\right].
\]

Thereby, if we define a critical gain $\bar{\rho}=\mbox{max}(\mbox{abs}({{\boldsymbol{\Lambda}}_{1}\matr{z}_{1}})),$
and consider a positive scalar constant $\rho_{0}$ , we obtain:
\[
\boldsymbol{\sigma}^{\footnotesize\mbox{T}}\dot{\boldsymbol{\sigma}}=-{\boldsymbol{\sigma}^{\footnotesize\mbox{T}}{\boldsymbol{\Lambda}}_{2}\boldsymbol{\sigma}-}\left[\begin{array}{c}
\sigma_{i}{\Lambda_{1}^{i,i}}z_{1}^{i}+(\bar{\rho}+\rho_{0})\left|\sigma_{i}\right|\\
\end{array}\right]<0,
\]
that ensures the invariance property to the block BSMC.

Note that the BSMC design above is related to a 2-block dynamic system (position/velocity). To apply it for a general $n$-block dynamic model (with a dynamic matrix in "lower triangular form"), just follow the ($n-1$) steps of the regular block backstepping procedure, as in  \cite{Vie:19,FS:99}, and then apply the methodology proposed here for the final step (n).

Now, we return to the proposed BSMC approach for the airship control case, to derive the final control law.
Once the sliding surface is defined as $ {\sigma} = {\matr{z}}_2 $, and its derivative is given by 
${\dot{\matr{z}}}_2 = -{\boldsymbol{\Lambda}_{1}}\matr{z}_{1}-{\boldsymbol{\Lambda}_{2}}\matr{z}_{2}-{\boldsymbol{\Lambda}_{s}}\sgn({\matr{z}_{2}})$ from \eqref{eq:LyapGeral},
then the proposed acceleration command $\ddot{\boldsymbol{\eta}}$ will be given by:
\[
 \ddot{\boldsymbol{\eta}}=-\boldsymbol{\Lambda}_{1}\matr{z}_1-\matr{K}_1\dot{\matr{z}}_1-\boldsymbol{\Lambda}_{2}\matr{z}_2-\boldsymbol{\Lambda}_{s}\sgn(\matr{z}_2)
\]
where $\boldsymbol{\Lambda}_{s}$ is a positive definite diagonal matrix, generalizing the scalar switching gain $\rho$, used previously.

Look that, like in the pure backstepping case, the term $\boldsymbol{\Lambda}_{1}\matr{z}_1$ is present in the control acceleration, being responsible for the independence in the gain adjustment and the performance enhancement, as we will see later.

For the evaluation of the airship control law, first recall that from Eq.~\eqref{eq:final:aceleracao}, we have the relations:

\begin{equation}
\begin{array}{l}
\matr{f} = \matr{M}(\dot{\matr{x}} - \matr{K}\matr{x}) -\matr{E}_{g}\matr{S}\matr{g}-\matr{F}_{a1}\\
\ddot{\boldsymbol{\eta}} = \matr{T}\dot{\matr{x}}+\matr{D}\boldsymbol{\Omega}_7\matr{C}\matr{x}
\implies 
\dot{\matr{x}}= \matr{T}^{+}(\ddot{\boldsymbol{\eta}} -\matr{D}\boldsymbol{\Omega}_7\matr{C}\matr{x}) \implies \\  \matr{M}\dot{\matr{x}}=\matr{M}\matr{T}^{+}(\ddot{\boldsymbol{\eta}} -\matr{D}\boldsymbol{\Omega}_7\matr{C}\matr{x})
\end{array}
\label{eq:final:aceleracao2}
\end{equation}

Such that we have the control law (input force) defined as:

\begin{equation}
\begin{array}{l}
\matr{f} = \matr{M}\matr{T}^{+}(-\boldsymbol{\Lambda}_1\matr{z}_1 - \matr{K}_1\dot{\matr{z}}_1 - \boldsymbol{\Lambda}_2\matr{z}_2
-\boldsymbol{\Lambda}_{s} \sgn{(\matr{z}_2)}
-\matr{D}\boldsymbol{\Omega}_7\matr{C}\matr{x}) \\
- \matr{M}\matr{K}\matr{x} -\matr{E}_{g}\matr{S}\matr{g}-\matr{F}_{a1}
\end{array}
\label{eq:imple:13}
\end{equation}

And using the definition of $ \matr{z}_2 $  from Eq.~\eqref{eq:imple:4}, as well as 
$ \matr{z}_1=\tilde{\boldsymbol{\eta}} $ and its derivative 
$ \dot{\matr{z}}_1= \dot{\tilde{\boldsymbol{\eta}}}=\matr{T}\matr{x}+\matr{B}\matr{v}_{w}-\dot{{\boldsymbol{\eta}}}_d$ from Eq.~\eqref{eq:final:aceleracao}, we come to the final BSMC control law as:
\begin{equation}
\begin{array}{l}
\matr{f} =
\matr{M}\matr{T}^{+}[-(\boldsymbol{\Lambda}_1+\boldsymbol{\Lambda}_2\matr{K}_1)\tilde{\boldsymbol{\eta}} - (\matr{K}_1+\boldsymbol{\Lambda}_2)\dot{\tilde{\boldsymbol{\eta}}}-\\
-\boldsymbol{\Lambda}_{s}
\sgn({\color{black}\dot{\widetilde{\boldsymbol{\eta}}}+\matr{K}_{1}\widetilde{\boldsymbol{\eta}}})-
-\matr{D}\boldsymbol{\Omega}_7\matr{C}\matr{x}]
- \matr{M}\matr{K}\matr{x} -\matr{E}_{g}\matr{S}\matr{g}-\matr{F}_{a1}
\end{array}
\end{equation}

Note that we consider here a wind speed estimation $\hat{\matr{v}}_{w}$
obtained from a wind estimator (shown in \cite{moutinho2016airship}), such that we have indeed $ \dot{\tilde{\boldsymbol{\eta}}}=\matr{T}\matr{x}+\matr{B}\hat{\matr{v}}_{w}-\dot{{\boldsymbol{\eta}}}_d$.
Also, in order to avoid the chattering in the control signal, instead of  $\sgn(\sigma)$ function, we indeed use the $\mbox{sat}(\sigma) $ function, defined as:
\begin{equation}
\mbox{sat}(\sigma)=\left({{\sigma}}/({|{\sigma}|+\epsilon})\right)
	\label{eq:suavizacao}
\end{equation} 
where $\epsilon>0$ is a constant boundary layer thickness.

Figure \ref{fig:principal:blocks} summarizes the MIMO BSMC design, showing the complete block diagram for the closed-loop system with airship model plus BSMC controller.

Look that to obtain the block diagram for a BS control design, just delete the gray sub-blocks from this figure, and to obtain the block diagram for the SMC control design, just delete the blue sub-block. This is the so called "unified framework design", presented in the next section.


%% file: part/controls/unified.tex
\subsection{Unified Framework for BS-SMC-BSMC Controllers}
\label{sec:unified}	

Now, regarding the MIMO BSMC project presented in the previous section, if we consider a sliding mode control design  (SMC) for the same airship model, where the slinding surface $\sigma$ is simply defined as the last z-error, or
$\sigma\triangleq \matr{z}_2$, then we come to the resulting SMC control laws and Lyapunov functions presented in Tables 1,2,3 below.
And if we consider a pure backstepping control design (BS), that is exactly the same BSMC design presented in the previous section, excluding the switching control term, then we come to the BS control presented in the same tables.

Thus, regarding the table of the formation law (Tab. 1), we note that both BS and SMC approaches differ from  the BSMC in a single term. For BS, this is the single switching term ${\boldsymbol{\Lambda}_{s}\sgn(\mathbf{z}_{2})}$, and for SMC this is the feedback term related to the primary error variable, or ${\boldsymbol{\Lambda}_{1}\mathbf{z}_{1}}$, recalling that the switching term increases the robustness to perturbations and modelling uncertainties, while the feedback of the primary error improves the performance of the time responses. 



\renewcommand*{\arraystretch}{1.2}
\begin{table}[!ht]
    \caption{\label{tab:table1} Lyapunov function, Lyapunov derivative and control law formation.}
	\centering
	\begin{tabular}{cc}
		\hline  
		 
		 \tabularnewline
		\hline
			& $\scriptstyle 
			\textcolor{blue}{\frac{1}{2}\mathbf{z}_{1}^{T}\boldsymbol{\Lambda}_{1}\mathbf{z}_{1}}+\textcolor{red}{\frac{1}{2}\mathbf{z}_{2}^{T}\mathbf{z}_{2}}$\\
		{\small BS} & $\scriptstyle 
		\textcolor{blue}{-\mathbf{z}_{1}^{T}\boldsymbol{\Lambda}_{1}\mathbf{K}_{1}\mathbf{z}_{1}}-\textcolor{red}{\mathbf{z}_{2}^{T}\boldsymbol{\Lambda}_{2}\mathbf{z}_{2}}<0$ \\
		&
		$\scriptstyle 
		\dot{\mathbf{z}}_2=-\textcolor{red}{\boldsymbol{\Lambda}_{2}\mathbf{z}_{2}}-\textcolor{blue}{\boldsymbol{\Lambda}_{1}\mathbf{z}_{1}}$
		\tabularnewline 
			\hline
			& $\scriptstyle 
			\textcolor{blue}{\frac{1}{2}\mathbf{z}_{1}^{T}\boldsymbol{\Lambda}_{1}\mathbf{z}_{1}}+\textcolor{red}{\frac{1}{2}\mathbf{z}_{2}^{T}\mathbf{z}_{2}}$\\
		 {\small BSMC}  &  $\scriptstyle 
		 \textcolor{blue}{-\mathbf{z}_{1}^{T}\boldsymbol{\Lambda}_{1}\mathbf{K}_{1}\mathbf{z}_{1}}-\textcolor{red}{\mathbf{z}_{2}^{T}\boldsymbol{\Lambda}_{2}\mathbf{z}_{2}}-\textcolor{gray}{\mathbf{z}_{2}^{T}\boldsymbol{\Lambda}_{s}\sgn(\mathbf{z}_{2})}<0$ \\ &  $\scriptstyle 
		 \dot{\mathbf{z}}_{2}=-\textcolor{red}{\boldsymbol{\Lambda}_{2}\mathbf{z}_{2}}-\textcolor{gray}{\boldsymbol{\Lambda}_{s}\sgn(\mathbf{z}_{2})}-\textcolor{blue}{\boldsymbol{\Lambda}_{1}\mathbf{z}_{1}}$
		\tabularnewline 
			\hline
		 	& $\scriptstyle 
		 	\textcolor{red}{\frac{1}{2}\mathbf{z}_{2}^{T}\mathbf{z}_{2}}$ \\
		{\small SMC} & $\scriptstyle 
		\textcolor{red}{\mathbf{z}_{2}^{T}\boldsymbol{\Lambda}_{2}\mathbf{z}_{2}}-\textcolor{gray}{\mathbf{z}_{2}^{T}\boldsymbol{\Lambda}_{s}\sgn(\mathbf{z}_{2})}<0$\\ 
			& $\scriptstyle 
			\dot{\mathbf{z}}_{2}=-\textcolor{red}{\boldsymbol{\Lambda}_{2}\mathbf{z}_{2}}-\textcolor{gray}{\boldsymbol{\Lambda}_{s}\sgn(\mathbf{z}_{2})}$
			\tabularnewline 
		\hline
	\end{tabular}  
\end{table}


\begingroup
\renewcommand*{\arraystretch}{1.2}
\begin{table}[!ht]
    \caption{\label{tab:table2} Final airship control law (force input vector) with $F_{ff}=\mathbf{M}\mathbf{K}\mathbf{x}+\mathbf{E}_{g}\mathbf{S}\mathbf{g}+\mathbf{F}_{a1}$.}
	\centering
	\begin{tabular}{cc}
		\hline
		 & {\small Control Law}
		\tabularnewline 
		\hline
		 {\small BS}	& 
		$ \begin{array}{c} \scriptstyle 
			\mathbf{f}=\mathbf{M}\mathbf{T}^{+}[-(\textcolor{red}{\boldsymbol{\Lambda}_{2}}+\mathbf{K}_{1})(\dot{\boldsymbol{\eta}}-\dot{\boldsymbol{\eta}}_{d})- \\
			\scriptstyle
			-(\textcolor{red}{\boldsymbol{\Lambda}_{2}}\mathbf{K}_{1}+\textcolor{blue}{\boldsymbol{\Lambda}_{1}})(\boldsymbol{\eta}-\boldsymbol{\eta}_{d})-\mathbf{D}\boldsymbol{\Omega}_{7}\mathbf{C}\mathbf{x}]-F_{ff}
			\end{array}$
		\tabularnewline 
		\hline
		{\small BSMC} 	&  
		$\begin{array}{c} \scriptstyle
		\mathbf{f}=\mathbf{M}\mathbf{T}^{+}[-(\textcolor{red}{\boldsymbol{\Lambda}_{2}}+\mathbf{K}_{1})(\dot{\boldsymbol{\eta}}-\dot{\boldsymbol{\eta}}_{d})- \\ \scriptstyle
		-(\textcolor{red}{\boldsymbol{\Lambda}_{2}}\mathbf{K}_{1}+\textcolor{blue}{\boldsymbol{\Lambda}_{1}})\left(\boldsymbol{\eta}-\boldsymbol{\eta}_{d}\right)-\textcolor{gray}{\boldsymbol{\Lambda}_{s}
		\sgn({\color{gray}\dot{\widetilde{\boldsymbol{\eta}}}+\matr{K}_{1}\widetilde{\boldsymbol{\eta}}})
		}
		-\mathbf{D}\boldsymbol{\Omega}_{7}\mathbf{C}\mathbf{x}]-F_{ff}
		\end{array}$
		\tabularnewline
		\hline
		{\small SMC} 	& $\begin{array}{c} \scriptstyle
		\mathbf{f}=\mathbf{M}\mathbf{T}^{+}[-(\textcolor{red}{\boldsymbol{\Lambda}_{2}}+\mathbf{K}_{1})(\dot{\boldsymbol{\eta}}-\dot{\boldsymbol{\eta}}_{d})- \\ \scriptstyle
		-(\textcolor{red}{\boldsymbol{\Lambda}_{2}}\mathbf{K}_{1})\left(\boldsymbol{\eta}-\boldsymbol{\eta}_{d}\right)-\textcolor{gray}{\boldsymbol{\Lambda}_{s}
		\sgn({\color{gray}\dot{\widetilde{\boldsymbol{\eta}}}+\matr{K}_{1}\widetilde{\boldsymbol{\eta}}})
		}
		-\mathbf{D}\boldsymbol{\Omega}_{7}\mathbf{C}\mathbf{x}]-F_{ff}
		\end{array}$
		\tabularnewline
		\hline
	\end{tabular}  
\end{table} 
\endgroup

\begingroup
\renewcommand*{\arraystretch}{1.2}
\begin{table}[!ht]
    \caption{\label{tab:table3} Control Law Gains.}
	\centering
	\begin{tabular}{cccc}
		\hline  
		 & {\small "P" gain}  & {\small "D" gain}
		 & {\small Switch gain} \tabularnewline
		\hline
		{\small BS} 	& \( \scriptstyle  -(\textcolor{red}{\boldsymbol{\Lambda}_{2}}\mathbf{K}_{1}+\textcolor{blue}{\boldsymbol{\Lambda}_{1}}) \)
		& \(\scriptstyle  -(\textcolor{red}{\boldsymbol{\Lambda}_{2}}+\mathbf{K}_{1})\)
		& \textendash
		\tabularnewline 
		{\small BSMC} 	& $ \scriptstyle -(\textcolor{red}{\boldsymbol{\Lambda}_{2}}\mathbf{K}_{1}+\textcolor{blue}{\boldsymbol{\Lambda}_{1}})$
		& $\scriptstyle  -(\textcolor{red}{\boldsymbol{\Lambda}_{2}}+\mathbf{K}_{1})$
		& $\scriptstyle  -\textcolor{gray}{\boldsymbol{\Lambda}_{s}}$
		\tabularnewline 
		{\small SMC} 	
		& $\scriptstyle  -(\textcolor{red}{\boldsymbol{\Lambda}_{2}}\mathbf{K}_{1})$
		& $\scriptstyle  -(\textcolor{red}{\boldsymbol{\Lambda}_{2}}+\mathbf{K}_{1})$
			& $\scriptstyle  -\textcolor{gray}{\boldsymbol{\Lambda}_{s}}$
		\tabularnewline 
		\hline
	\end{tabular}  
\end{table} 
\endgroup

Note that this unified framework basis is only possible due to the special selected sliding manifold $\sigma=\matr{z}_{2}$, used in the BSMC project (from BSMC-2), such that its performance, comparatively to BS and SMC, can be easily analysed and evaluated. In the case that the expressions for the three control laws of BS, SMC and BSMC are too different, like in the BSMC-1 project, then a theoretical-based comparative analysis is hard to be done. That is why simulation/experimental results are so commonly used, in the literature, as a tool for comparative performance analysis. In \cite{2013UAV}, for example, the authors compare BS, SMC, BSMC and BHSMC for a fixed wing UAV control, based on experimental tracking results.

Moreover, from the tables of control law (Tab. 2) and control gains (Tab. 3), we note that the ``proportional'' and ``derivative'' feedback gains of the BSMC approach can be independently adjusted through the matricial gain ${\boldsymbol{\Lambda}_{1}}$, like in the pure backstepping case \cite{FS:99}. This makes it easier, for BS and BSMC, to increase the proportional gain to reduce the steady state error, without affecting the derivative gain. Another interesting point is that this independence in gain tunning is only possible due to Property 1, regarding cross-coupling cancellation of the backstepping procedure (\ref{eq:prop1}), that is preserved when we use $\sigma=\matr{z}_{n}$. Also, if desired, a variable switching gain for the nonlinear control term can be used, in order to keep the  attraction/invariance of Property 2 (Eq. \ref{eq:prop2}).


%% file: results.tex
\section{Simulation Results and Analysis}

This section presents the simulation results, using the NOAMAY airship simulator, with the three nonlinear controllers (BS,SMC,BSMC) in the proposed unified framework. In the first two parts, we show the results for the positioning problem, with ideal and actual controllers, and in the third part we detail the results for a complete 3D mission.

In these test cases, all the controllers gains are  positive definite diagonal matrices of dimension $  \realnumber^{7\times7} $, with the first three elements of the diagonal corresponding to the position errors, while the last four elements are related to the angular errors given by the quaternions. And, except when otherwise stated, we use the controller gains below in the simulations:
\begin{equation}
\begin{array}{l}
\scriptstyle
\matr{K}_1 = 0.2\matr{I}_7 ~~~~~~ 
\boldsymbol{\Lambda}_1 = \mbox{diag}[ 0.05,0.05,0.05,0.2,0.2,0.2,0.2
] \\
\scriptstyle
\boldsymbol{\Lambda}_2 = 0.5\matr{I}_7 ~~~~~~ 
\boldsymbol{\Lambda}_s = \mbox{diag}[
0.1,0.1,0.1,0.2,0.2,0.2,0.2
] \,\,\,\,\,\,\,\,\,\,\,\,   { \epsilon}=0.1
\end{array}
\label{ganha:padrao}
\end{equation}

Note that we use here a fixed swithing gain ${\Lambda}_s$ in the simulations, such that we do not consider the imposition of the reaching/invariance (Property 1), and the dual behavior of the BSMC controller may occur. 

\subsection{Positioning Problem - Ideal Controllers}



In this first part, we consider the positioning problem for the hovering flight over a fixed target in the "ideal" condition. This means that the applied control action is not subject to saturations or delays (either dynamic or dead time). Also, the switching term in the SMC and BSMC controllers comes from a pure signal function, without chattering smoothing. This is important to further evaluate which kind of basic features or properties of the controllers may be lost when used within the real limitations of a real actuator. Also, the "ideal" controller simulations clearly illustrate the dual behavior of the proposed BSMC controller.

\begin{figure}[!htbp]
\centering
\includegraphics[scale=0.25]{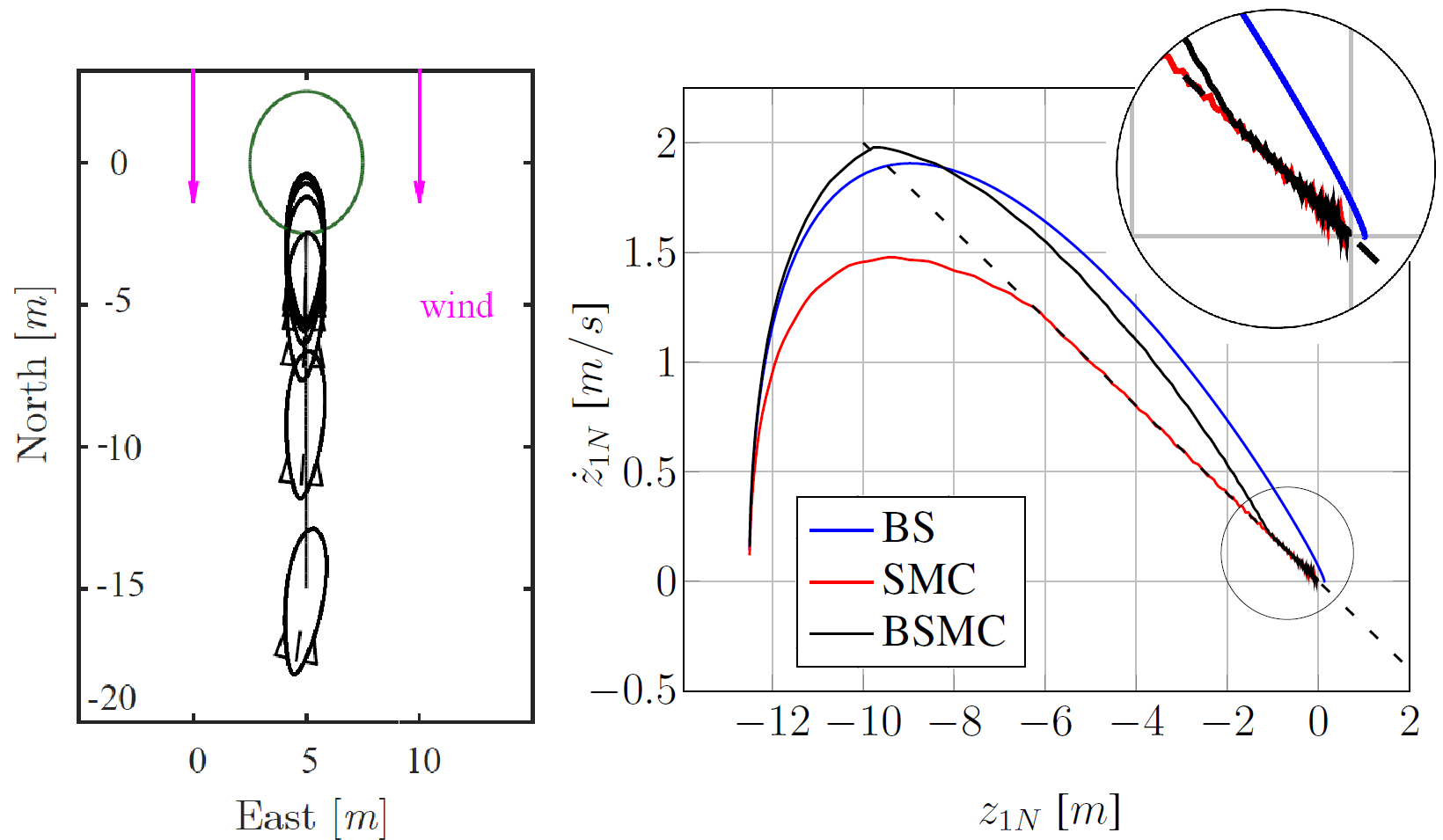}
\caption{Left: positioning with BSMC. Right: phase plane for $z_1 \times \dot{z}_1$, ~$z_1 = North$ error. Ideal controllers case.\label{fig:ideal:trajetoria}}
\end{figure}

The test simulation for the hovering flight starts with the airship initially hovering over a given ground point ($N=-15m,~E=0$), at a given altitude ($D=-50 m$), and then it starts to move toward a target located 15 meters ahead (in North direction), at the same altitude, where the airship should again stay hovering (Figure~\ref{fig:ideal:trajetoria}). To allow the final heading of the airship against the wind direction, which is a necessary condition for its stability, we consider a virtual circular target of radius $2.5$ meters around the true point of interest, in order to assure a one-degree freedom in the final position \cite{azinheira2006airship}. Also, we assume that the airship starts the mission slightly misaligned in $10^\circ $ in yaw, pitch and roll, and there is a wind perturbation of $ 1m/s $ blowing from the North (Fig.~\ref{fig:ideal:trajetoria}).

Figure ~\ref{fig:ideal:trajetoria}-left shows the resulting trajectory of the  airship for the BSMC controller case, and Fig~\ref{fig:ideal:trajetoria}-right shows the phase plane of the corresponding North Position $\times$ North Velocity for the three controllers (BS, SMC and BSMC). 
This 2D phase portrait allows for a clear evaluation of the dynamic behavior of the closed-loop systems. First,  note that the sliding surface is given by the linear relation $\matr{z}_2=\dot{\matr{z}}_1+\matr{K}_1\matr{z}_1=0$ in the phase plane.
Thus, we see that for the SMC controller, the state error  approaches, in finite time, this sliding surface ($\matr{z}_2=0$),  and then slides asymptotically toward the origin, when $\matr{z}_1=\matr{0} $ (look chattering in zoom). The error of the BS controller case evolves like a typical stable node atraction in the state space. And, finally, the state error for the BSMC controller evolves initially like a stable node, but later it is attracted by the sliding surface, sliding toward the origin.



\begin{figure}[!htbp]
\centering
\includegraphics[scale=.21]{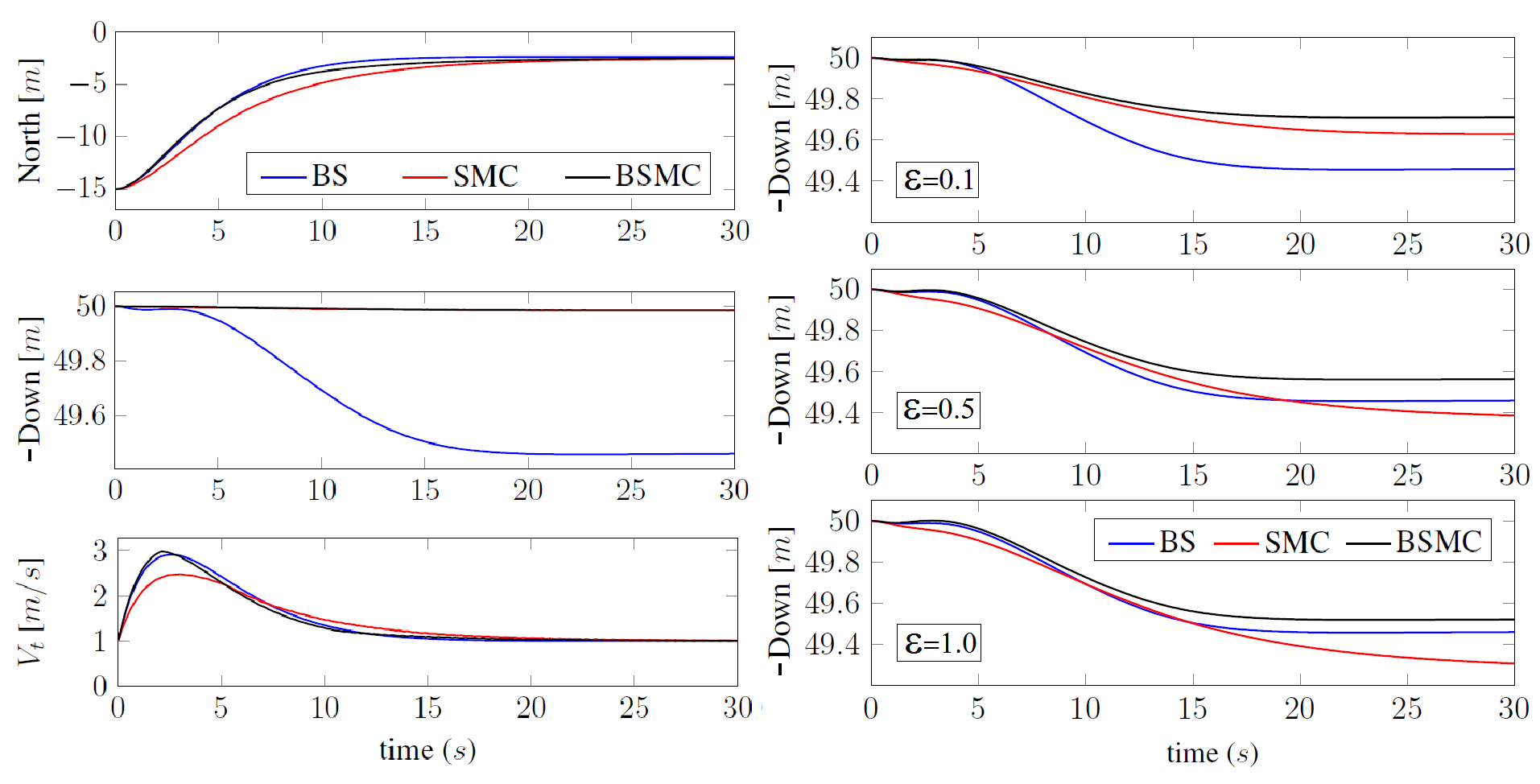}
\caption{Left: $N,-D, V_T$. Right: Altitude for different $ \epsilon $. Ideal controllers.
\label{fig:ideal:u}}
\end{figure}


Figure~\ref{fig:ideal:u} shows the corresponding evolution of the North and -Down (altitude) positions, as well as the true airspeed signal, which is equal to $V_t=1 m/s$ at the initial and final instant of times due to the heading wind. Note that the BS and BSMC responses are slightly faster than those of SMC. Also, there is an offset error in altitude for the BS, not seen in the ideal SMC and BSMC due to the switching term compensation.
\begin{figure}[!htbp]
\centering
\includegraphics[scale=.26]{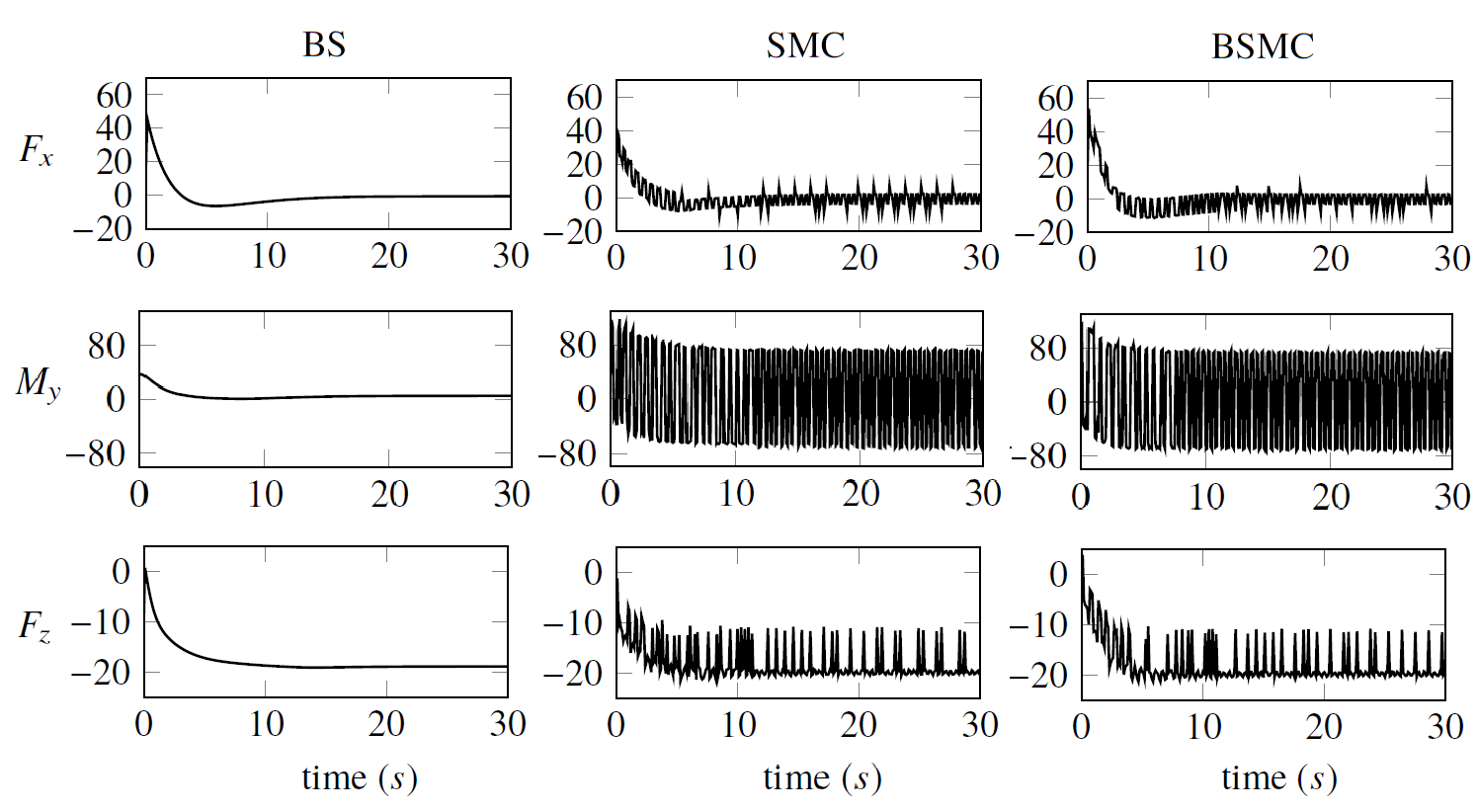}\\
\caption{Longitudinal actuators signals $(F_x , M_y $ and $ F_z $). Ideal controllers.\label{fig:ideal:forca}}
\end{figure}


Figure~\ref{fig:ideal:forca} shows the longitudinal force/moment control signals $ F_x $, $ M_y $ and $ F_z $.
Note that the response time is slightly shorter, in the BSMC case for the three control inputs $F_x $, $ M_y $ and $ F_z $, probably due to the  additional term in the proportional gain (with $\boldsymbol{\Lambda}_1$). 

\begin{figure}[!htbp]
\centering
\includegraphics[scale=.26]{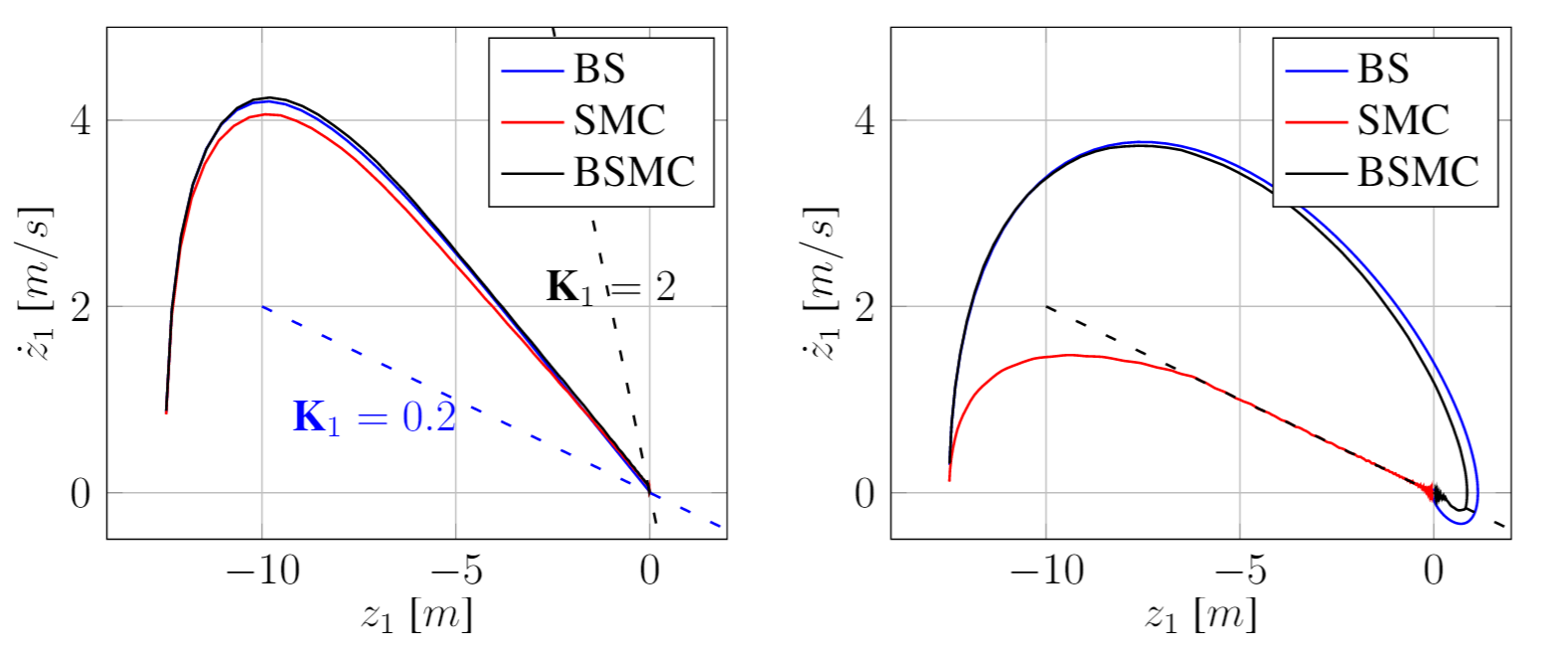}\\
\caption{Phase planes for North errors (Position $ \times $ Velocity), with $ \matr{K}_1 = 2.0\matr{I}_7 $,~~~ $\boldsymbol{\Lambda}_2 = 0.5\matr{I}_7,~~ \boldsymbol{\Lambda}_s = \text{diag}([0.1~0.1~0.1~0.2~0.2~0.2~0.2])$, and also (a)~
 $ \boldsymbol{\Lambda}_1 = \text{diag}([0.05~0.05~0.05~0.2~0.2~0.2~0.2])$; (b) $ \boldsymbol{\Lambda}_1 = \text{diag}([0.25~0.25~0.25~0.2~0.2~0.2~0.2])$.  
\label{fig:increased}}
\end{figure}

To see the influence of $ \matr{K}_1 $ and $\boldsymbol{\Lambda}_1$ gains in the dynamics of the closed-loop with the different controllers, we simulated this same example with  increased values of $ \matr{K}_1 $ and $\boldsymbol{\Lambda}_1$, according to Fig. \ref{fig:increased}. Note that with an increased  $ \matr{K}_1 $ gain, the  SMC controller tends to behave like a BS or BSMC, as this gain modifies the slope of the sliding surface, with the final  error path changed according to the system physical properties. 

Now, if we add a smoothing action in the switching controller terms of SMC and BSMC, in order to reduce the chattering, recalling the definition of $\epsilon$, from Eq. (\ref{eq:suavizacao}), we can make some interesting observations. The results are shown in Figure  \ref{fig:ideal:u}-right for the -Down position error, considering $\epsilon=0.1,$ $0.5$ and $1.0$. As expected, the greater the $\epsilon$ value, the less will be the chattering in the control signal, with corresponding greater offset errors in the controlled variables. This can be confirmed in the -Down signal shown in Fig. \ref{fig:ideal:u} that should stay at the reference value of $ 50 m $ (desired altitude). We can see that the performance is degraded for higher $\epsilon$ values in the SMC case, becoming even worse than BS case for $\epsilon=0.5$ and $1.0$. However, it is interesting to note that the same does not happen for the BSMC controller, where the degradation is mitigated. This is due to the fact that the "proportional" controller gain for BS and BSMC cases are higher than that of SMC due to the $\boldsymbol{\Lambda}_1$ term. Thus, we see that the  position  feedback error gain, coming from the backstepping part of BSMC helps to compensate for the loss of accuracy in the  switching term coming from SMC.

\subsection{Positioning Problem - Actual Controllers}

In this subsection, we add constraints and limitations to the control law, in order to simulate a real actuator condition for a true application. Firstly, we consider the use of the chattering suavization parameter $\epsilon$, from Equation  (\ref{eq:suavizacao}). Second, we impose saturation limits on the forces and moments as $ F_x = \pm 150 N $, $ F_y = \pm 30 N $, $ F_z = \pm 35 N $, $ M_x = \pm 10 Nm $, $ M_y = \pm 90 Nm $ and $ M_z = \pm 90 Nm $, according to the guidelines suggested in our previous work of \cite{AM:08}. Note that 
$ F_y$ and $F_z$ are subject to more stringent constraints, as the tilting of the 4 electrical propellers is mainly devoted to yield longitudinal force components. Finally, we add a first-order dynamics at the force output signal of the controllers, with a time constant of $ 0.5 $ sec, in order to simulate the influence of the dynamics of the  actuators (motors and tail surfaces).

\begin{figure}[!htbp]
\centering
\includegraphics[scale=.30]{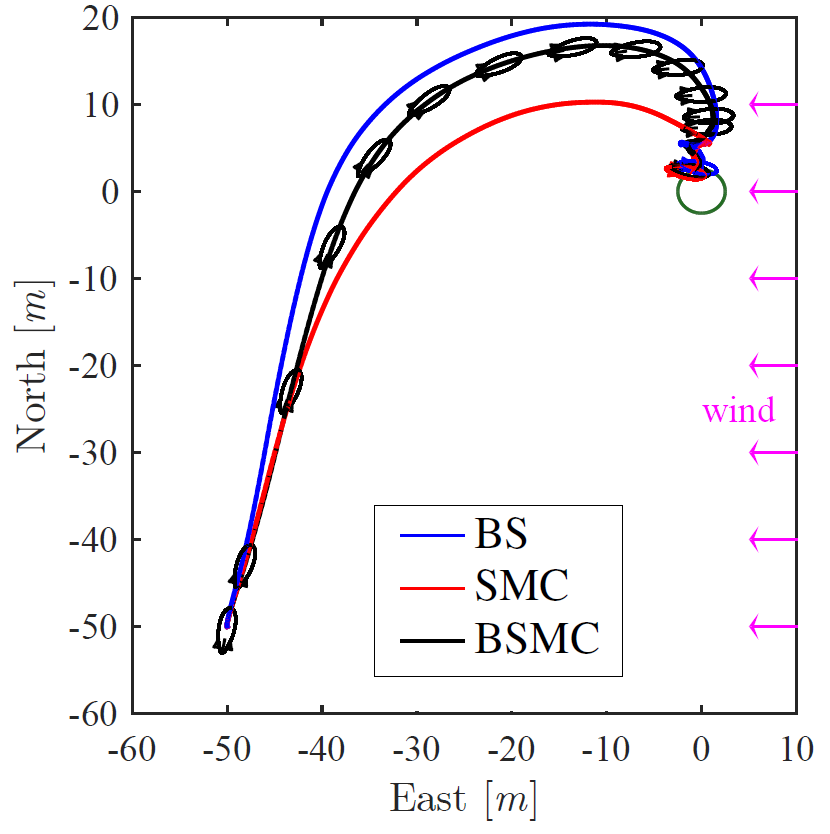}
\caption{Hovering flight path in $East \times North$. Actual controllers. \label{fig:pairado:pos:ned1}}
\end{figure}

\begin{figure}[!htbp]
\centering
\includegraphics[scale=.26]{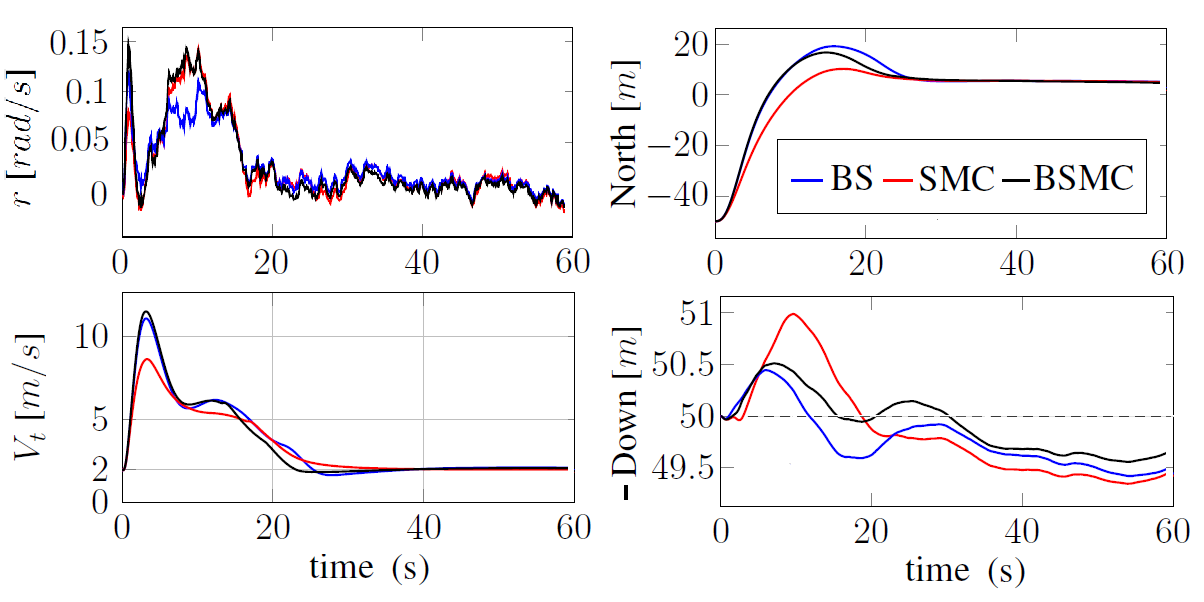}
\caption{Hovering flight: yaw rate $r$, aispeed  $V_t$, positions $N, -D$.\label{fig:pairado:pos:ned2}}
\end{figure}

The positioning control test, proposed here, has the same flight configuration found in the work of \cite{YWZ:16}. The initial position is given by $ (N,E,D)=(- 50, ~ -50, ~ 50) $ and the final position is $ (N,E,D)=(0,0, ~ 50) $ with a tolerance radius of $ 2.5m $ around the target. A wind estimator with a Kalman filter is used in the simulations as presented in \cite{moutinho2016airship}, based on a combination of both GPS and Pitot tube sensors. The wind speed is $ 2m / s $ with incidence angle of $90^{\circ}$ plus a turbulence of $ 1m / s $ of standard deviation, as shown in Figure~\ref{fig:pairado:pos:ned1}. This figure shows the airship trajectory in $North \times East$, and Figure~\ref{fig:pairado:pos:ned2} shows the time responses for $North$, $-Down$, $r$ and airspeed $V_t$, for the three nonlinear controllers.

From the North-East path, we note that BS and BSMC controllers yield a more aggressive response than the SMC case, with higher airspeeds and with an oveshoot in North direction, due to the extra $\boldsymbol{\Lambda}_1$  gain in the "proportional" term. From Figure~\ref{fig:pairado:pos:ned2}, it is interesting to note that the overshoot in North is bigger (from the path plot) and longer (from the North plot) for the BS controller, as compared with the overshoot of BSMC. And as the only difference between both controllers is the nonlinear switching term (from Tables 2 and 3 of section III.C), we can conclude that this term, still with diminished efficiency due to the actual controller constraints, is the responsible for a better performance in the positioning control task for BSMC. And when we compare SMC with BSMC, we see that this last one generates a slightly faster response, due to the  $\boldsymbol{\Lambda}_1$ gain. The final altitude shows a steady-state error for all controllers, due to the turbulence and the wind estimator. However, this error is bigger for the SMC case, and smaller for the BSMC case.

\begin{figure}[!htbp]
\centering
\includegraphics[scale=0.30]{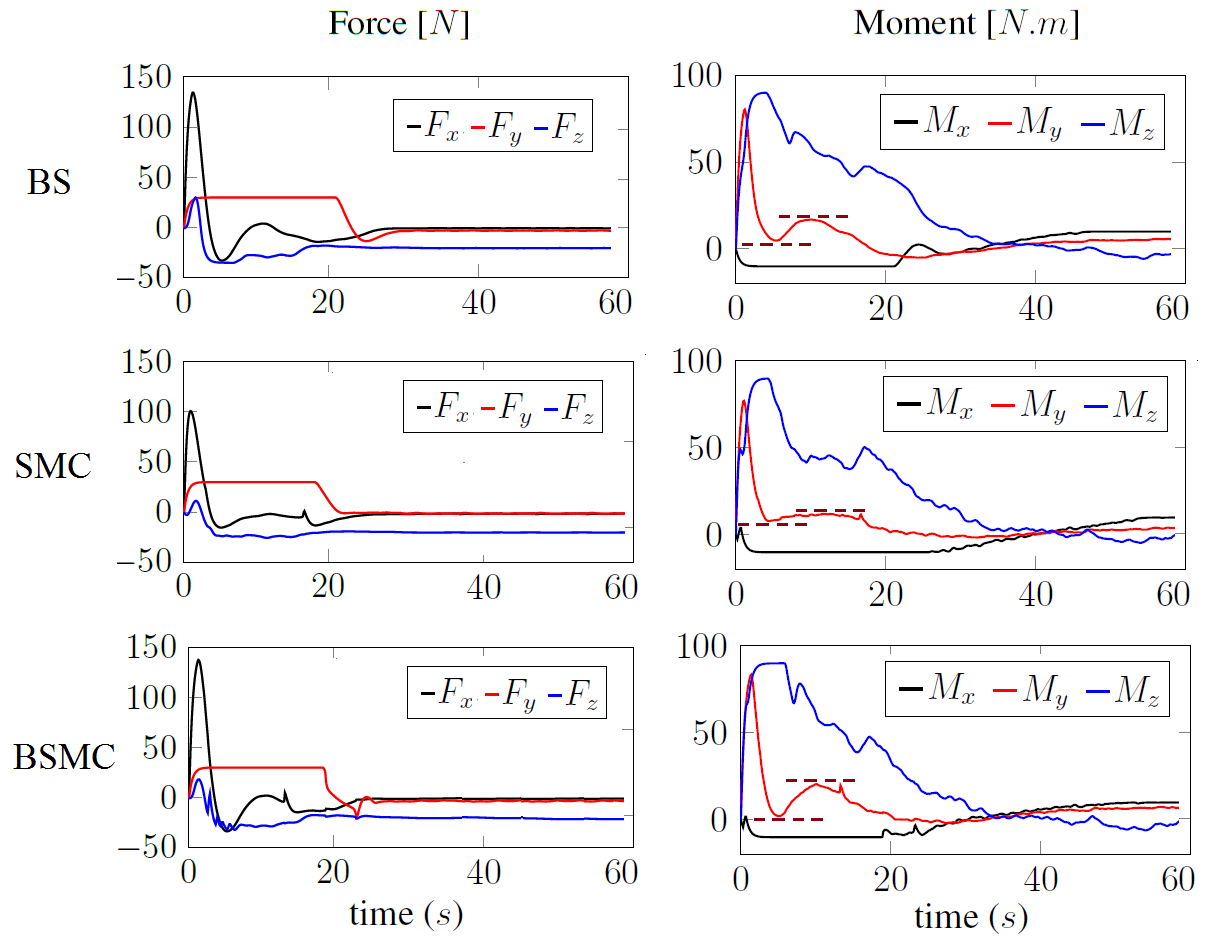}
\caption{Control force/moment signals for hovering flight.Actual controllers.  \label{fig:pairado:FM:bs}}
\end{figure}

The corresponding forces and moments for this test case are shown in Figure \ref{fig:pairado:FM:bs}. We can see the long saturation time in the lateral force for all nonlinear controllers, due to the reduced airship lateral actuation. The control signals for the BSMC case show a faster response, slightly undamped, and a short saturation in the yaw moment. One interesting analysis can be done, comparing the amplitude range of the pitch moment $M_y$, between $t=0$ and $t=20s$, for the 3 controllers, as marked through horizontal dashed bars in this figure. We can see that the amplitude change in this control signal is bigger for BSMC when compared to both SMC and BS. This stronger corrective action may be responsible for the smaller errors in altitude for BSMC controller, as the pitch moment is closely related to the vertical movement. Moreover, look that, in the first 10 seconds of flight, the vertical force $F_z$ is much more active for BSMC than for SMC (due to $\boldsymbol{\Lambda}_1$), which may explain the increased altitude error for this last one.
In addition, the yaw moment $M_z$, in the time interval of $t=0-20$, shows a sustained higher value for BSMC when compared to BS (due to the switching term), resulting in a higher yaw rate around $t=10s$ (Figure \ref{fig:pairado:pos:ned2}), which explains the higher North overshoot for the BS controller. 


Also, when we compare the altitude errors (-Down signal) in figure \ref{fig:ideal:u} (ideal controllers) and figure \ref{fig:pairado:pos:ned2} (actual controllers),
we see that in the first case, the switching term is fundamental 
to yield a null altitude error for BSMC as compared to BS. However, in the second case, both the switching term and the $\boldsymbol{\Lambda}_1$ gain help to reduce the altitude error, contributing for a performance improvement with the BSMC approach.

As cited before, this same example was treated in \cite{YWZ:16}, where the authors compare SMC and BSMC controllers for the positioning problem of a bigger airship. However, a detailed comparison is not possible as, in their work, the authors did not impose saturation constraints on the control forces, even for the lateral forces. As a result of this assumption, in their case there is no overshoot in the resulting North path as observed in our case. Another interesting point is that they use a type-1 BSMC approach, instead of the type-2 used here, such that they consider the sliding surface $\sigma=\matr{K}_1\mathbf{z}_{1}+\mathbf{z}_{2}$ instead of $\sigma=\mathbf{z}_{2}$ considered here. This means that, if on one hand, their BSMC does not present a dual behavior (as the sliding surface is always attractive), on the other hand, their control law does not have the independence in the gain adjust, as the term $\boldsymbol{\Lambda}_1$ in the "proportional" gain is not present.

\begin{figure}[!htbp]
\centering
\includegraphics[scale=0.26]{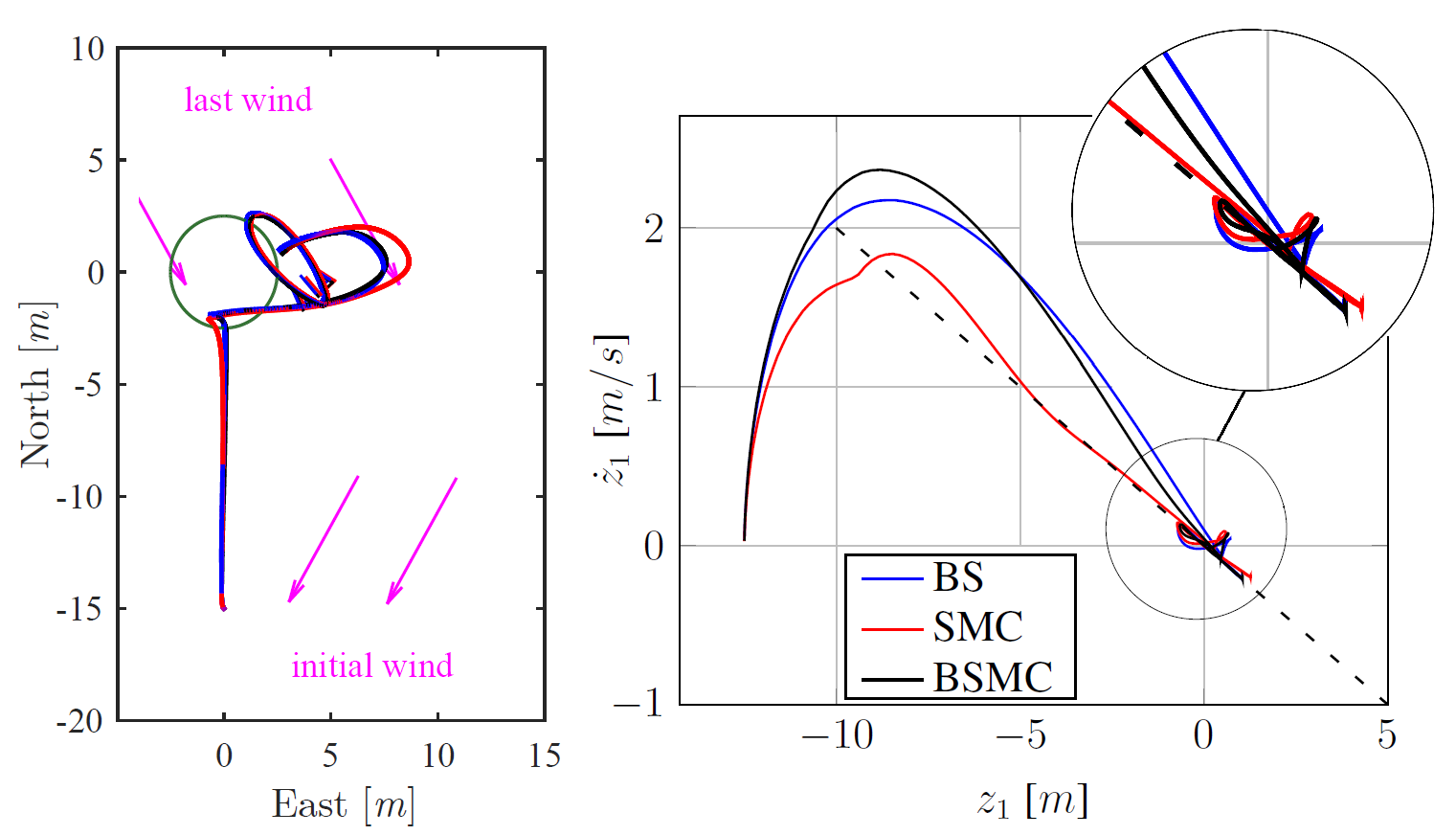}
\caption{Left: hovering with windshift disturbance. Right: phase plot for North errors. Actual controllers case.}
\label{fig:pairado:real:traj}
\end{figure}

Finally, as a test case for robustness against external perturbations, we simulate the same hovering example, adding now a sudden windshift \cite{Vie:19}, while the airship is over the target. The path results are shown in Figure ~\ref{fig:pairado:real:traj}, together with the corresponding phase plane for the North errors. We can see that the 3 controllers are robust to reject the perturbation, but the SMC case shows a higher distance error during the repositioning phase. The phase plot also shows the effect of chattering reduction/smoothing, as the SMC shows a boundary layer in the sliding manifold, and the state errors slides on this surface only for small error values, when the switching term, with a fixed gain, has a higher influence.

\subsection{Path/Velocity Tracking in 3D Mission}

\begin{figure}[!htbp]
\centering
\includegraphics[scale=.24]{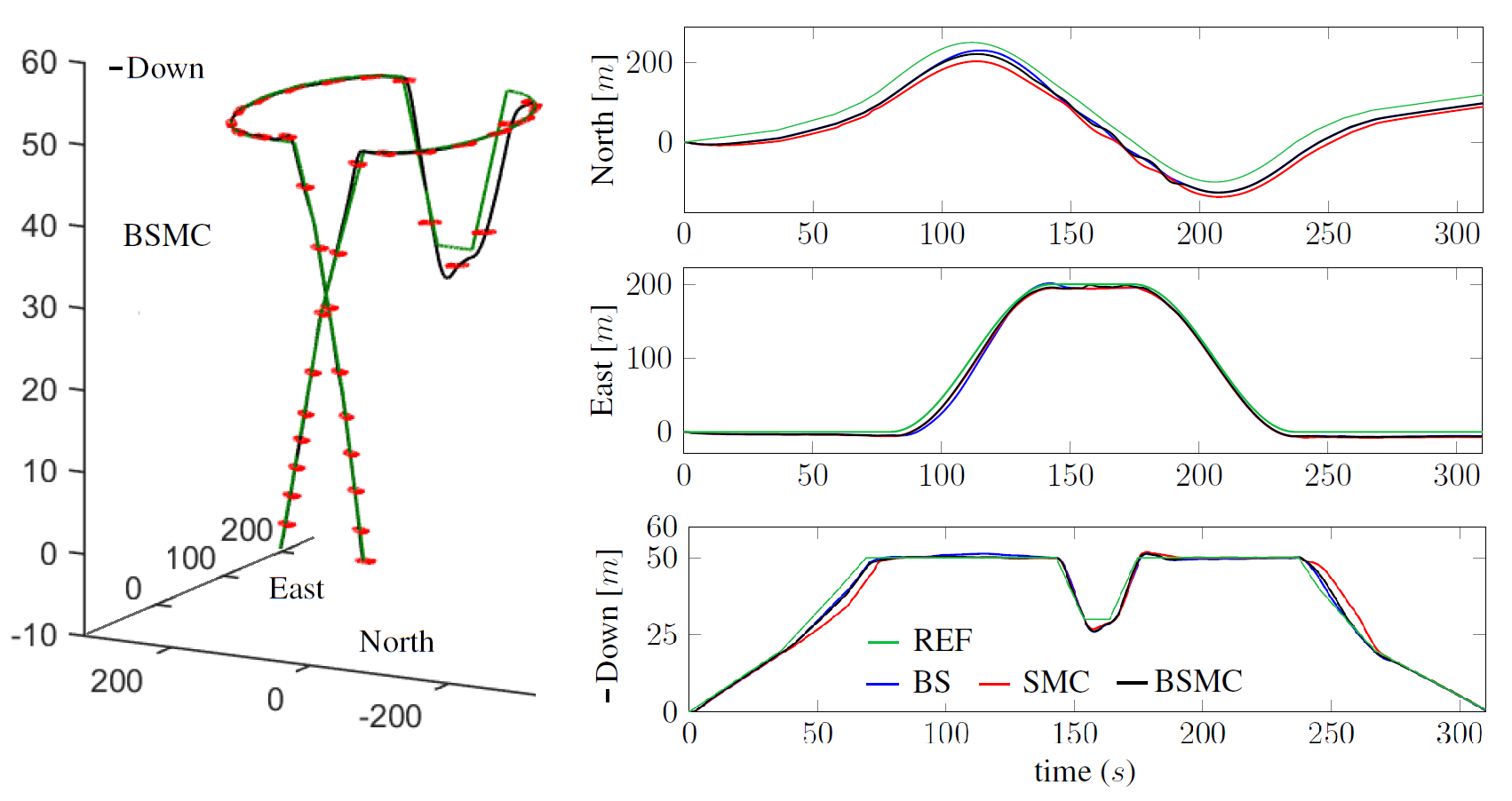}\\
\caption{Airship 3D trajectory and position errors for complete flight mission. \label{fig:per:ned1}}
\end{figure}

In this subsection, we present the simulation results for a complete 3D flight mission, with vertical take-off/landing, path tracking and velocity tracking, for the three control approaches. The resulting airship 3D trajectory is shown in Figure \ref{fig:per:ned1} for the BSMC controller only, as those from SMC and BSMC are very similar.
 The controller gains considered here  are the same presented previously in Eq. (\ref{ganha:padrao}). We consider the presence of a wind speed of $ 2m/s $ with $ 30^\circ $ incidence from North plus a turbulence of $ 1m/s $ std. dev.



\begin{figure}[!htbp]
\centering
\includegraphics[scale=.25]{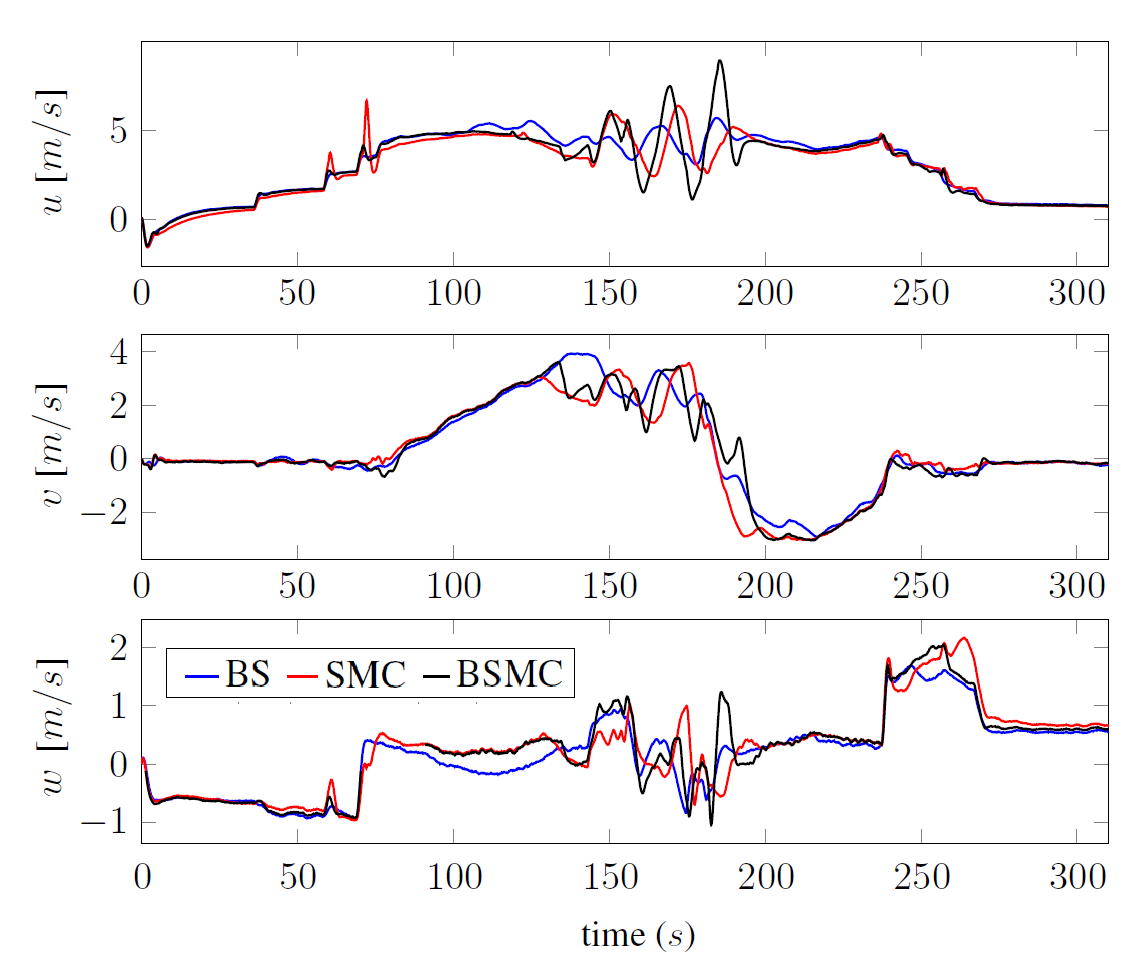}\\
\label{fig:per:uvw}
\caption{Linear velocities for the controllers in 3D mission.}
\label{fig:per:velocidade1}
\end{figure}


\begin{figure}[!ht]
\centering
\includegraphics[scale=.25]{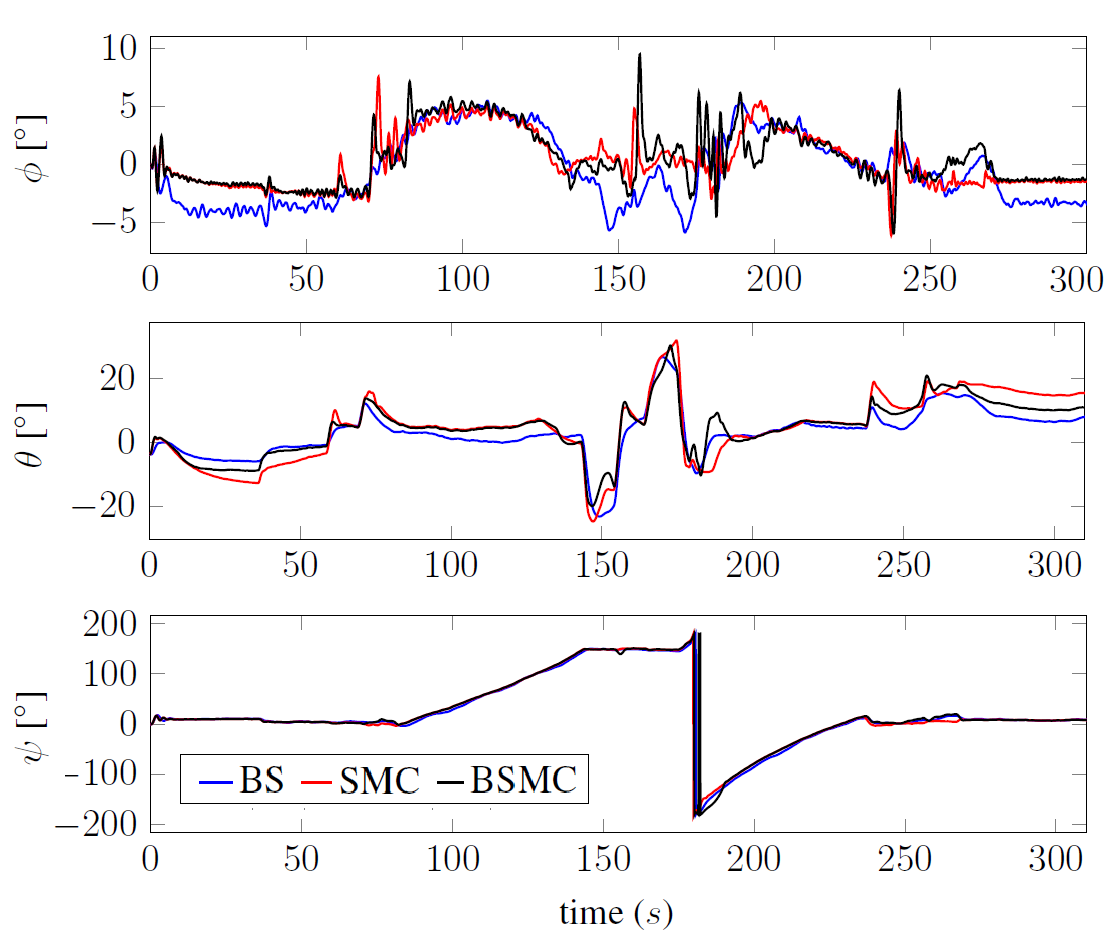}\\
\caption{Euler angles angles for 3D mission flight.\label{fig:per:angulo:velocidade1}}
\end{figure}

\begin{figure}[!ht]
\centering
\includegraphics[scale=.26]{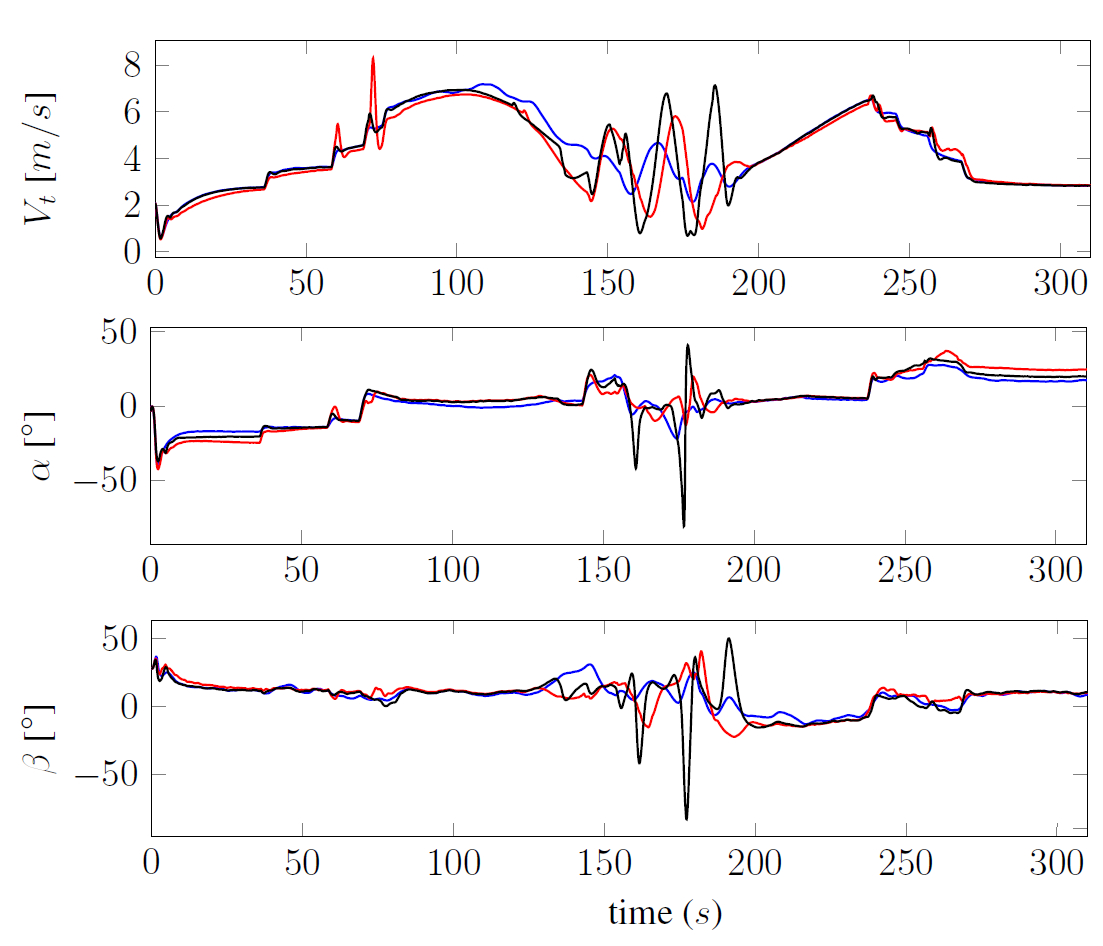}\\
\caption{Airspeed $V_t$ and aerodynamic angles for 3D mission flight.\label{fig:per:angulo:velocidade2}}
\end{figure}

\begin{figure}[!ht]
\centering
\includegraphics[scale=.22]{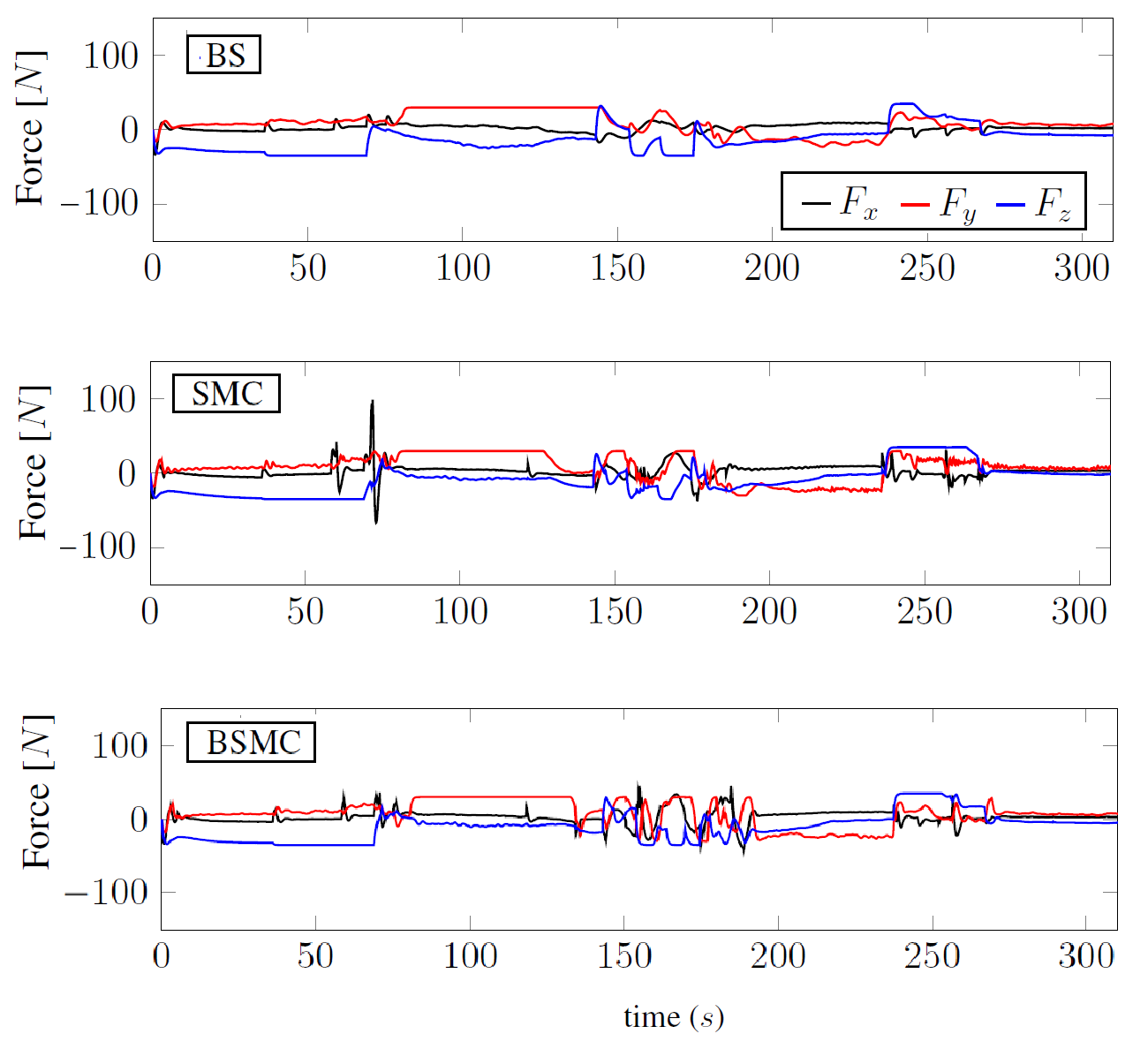}\\
\caption{Forces for the controllers in 3D mission flight.
\label{fig:per:forca1}}
\end{figure}

\begin{figure}[!ht]
\centering
\includegraphics[scale=.22]{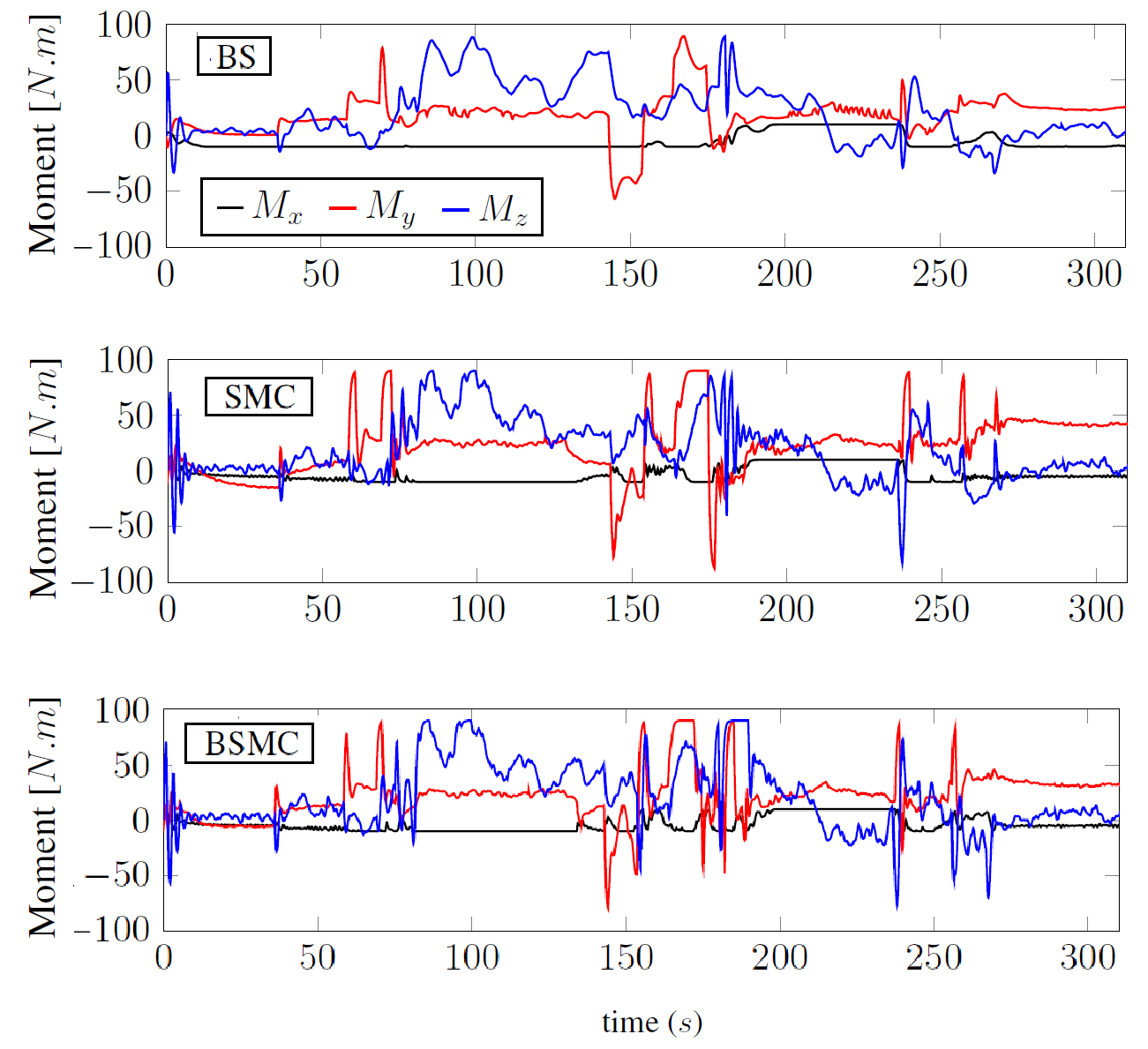}\\
\caption{Moments for the controllers in 3D mission flight. 
\label{fig:per:forca2}}
\end{figure}

Figure \ref{fig:per:ned1} also shows the references and the time responses for (N,E,D) positions for all the 3 controllers. Initially, the airship takes off almost vertically with a speed of about $w=1m/s$ (Figure \ref{fig:per:velocidade1}).
Then, it performs an ellipsoidal trajectory, at a constant altitude ($h=50m$) and constant groundspeed ($u=5m/s$). At about $t=150s$, the airship follows a change in altitude command, as observed in the figure, while keeping a constant East position. Then, it recovers the previous altitude, completing the ellipsoidal path, with a final vertical landing.

Figures~\ref{fig:per:velocidade1} shows the time responses for the linear airship velocities, where we can see a slightly oscillatory response (for SMC and BSMC) during the altitude shift, in the middle of the mission.

Figure~\ref{fig:per:angulo:velocidade1} shows the behavior of the Euler angles ($\phi$, $\theta$ and $\psi$), as well as the aerodynamic angles and the airspeed $V_t$.  As this is an oval trajectory, and the wind incidence angle is constant, the airship will be subject to a wide variation in the relative wind incidence, as we can see from the $V_t$ airspeed response (Figure ~\ref{fig:per:angulo:velocidade2}), that varies from about $1m/s$ to nearly $8m/s$ during the overall path. This challenging condition justifies the use of a nonlinear controller, as the airship dtynamics is strongly dependent on the airspeed, due to the reduced tail surfaces efficiency  at lower airspeeds.

Figures ~\ref{fig:per:forca1} and ~\ref{fig:per:forca2} show the forces and moments generated by each controller in this 3D flight mission. We can observe the saturations, mainly in the lateral force $F_y$ and the yaw moment $M_z$, but somehow on the vertical force $F_z$. The saturation in the lateral force, during the circular maneuvers, lasts longer for the BS controller and less for the SMC controller. 

The overall evaluation is the tracking performance seems similar for the three controllers, but the tracking errors, both in distance to the trajectory as well as the commanded velocities are slightly lower for the BSMC controller, despite its greater control effort. This is due to the presence of the extra controller term $\boldsymbol{\Lambda}_1$ in the "proportional" gain that helps to compensate for the degrading performance of the switching term when the control constraints are imposed.

%% file: conclusion.tex
\section{Conclusions}
\label{sec:Concl}	

The theoretical fundamentals of backstepping sliding mode control (BSMC) are scarce in the scientific literature, as there is a general sense that the features/performance of the combined approach "BS+SMC" are simply the sum of those of its original components, backstepping (BS) and sliding mode (SMC). However, we show here that when joining BS with SMC, some original properties of these two approaches may be lost in the combined BSMC formulation.

Firstly, we have derived a unified framework basis for the design/analysis of vectorial BSMC, that indeed can be also used for the design of the vectorial SMC and the vectorial BS. The framework basis can be applied to any model in a “lower triangular block” form, which is typical in mechanical systems, where the position and velocity vectors are naturally cascaded. 
The unified framework simplifies the comparative analysis, within a theoretical basis, between the three control approaches, allowing to highlight the advantages and drawbacks between them.
Second, the vectorial backstepping sliding mode control (BSMC), designed with this framework, was applied for the
positioning/tracking control of an autonomous airship, using a 6DOF simulator, showing a superior perfomance when compared to SMC and BS.
Within this unified formulation, we are able to identify the exact terms in the control laws that are responsible for a given feature/property observed in the simulation results.
Finally, we have not seen in the literature, any reference about the possible dual-behavior of a BSMC control, as illustrated here through simple phase planes.